\documentclass{aa}

\usepackage{graphicx}
\usepackage{txfonts}
\usepackage{natbib}
\usepackage{dcolumn}
\newcolumntype{.}{D{.}{.}{-1}}
\newcolumntype{;}{D{;}{.}{7}}
\bibpunct{(}{)}{;}{a}{}{,} % to follow the A&A style

\newcommand{\solm}{M$_{\odot}$\ }

\begin{document}

\authorrunning{D. Kunneriath, G. Witzel, A. Eckart, M. Zamaninasab et al. }
\titlerunning{Sgr$\,$A* Flares}
\title{Coordinated NIR/mm observations of flare emission from Sagittarius~A*}
\subtitle{}

\author{D. Kunneriath$^{1,2}$,
       G. Witzel$^1$,
       A. Eckart$^{1,2}$,
       M. Zamaninasab$^{2,1}$,
       R. Gie\ss \"ubel$^{2,1}$,
       R. Sch\"odel$^{3}$,
       F. K. Baganoff$^4$,
       M. R. Morris$^5$,
       M. Dov\v{c}iak$^{6}$, 
       W.J. Duschl$^{7,8}$,  
       M. Garc\'{\i}a-Mar\'{\i}n$^1$,
       V. Karas$^{6}$, 
       S. K\"onig$^{1}$,
       T. P. Krichbaum$^{2}$,
       M. Krips$^{12}$, 
       R.-S. Lu$^{2,1}$,
       J. Mauerhan$^{16}$,
       J. Moultaka$^{9}$,
       K. Mu\v{z}i\'c$^{1}$,
       N. Sabha$^{1}$,
       F. Najarro$^{10}$,
       J.-U. Pott$^{11}$,
       K. F. Schuster$^{12}$,
       L. O. Sjouwerman$^{13}$,
       C. Straubmeier$^1$,
       C. Thum$^{12}$,
       S. N. Vogel$^{14}$,
       P. Teuben$^{14}$,
       A. Weiss$^2$,
       H. Wiesemeyer$^{15}$,
       J. A. Zensus$^{2,1}$
}

\offprints{D. Kunneriath (devaky@ph1.uni-koeln.de)}

   \institute{ I.Physikalisches Institut, Universit\"at zu K\"oln,
              Z\"ulpicher Str.77, 50937 K\"oln, Germany
         \and
             Max-Planck-Institut f\"ur Radioastronomie, 
             Auf dem H\"ugel 69, 53121 Bonn, Germany
         \and
  Instituto de Astrof\'isica de Andaluc\'ia (CSIC), Camino Bajo de
    Hu\'etor 50, 18008 Granada, Spain 
         \and
 MKI, Massachusetts Institute of
           Technology, Cambridge, MA~02139-4307, USA 
         \and
  Department of Physics and Astronomy, University of California, 
     Los Angeles, CA 90095-1547, USA
         \and
 Astronomical Institute, Academy of Sciences, 
        Bo\v{c}n\'{i} II, CZ-14131 Prague, Czech Republic 
         \and
 Institut f\"ur Theoretische Physik und Astrophysik,
        Christian-Albrechts-Universit\"at zu Kiel, Leibnizstr. 15
        24118 Kiel, Germany 
         \and
 Steward Observatory, The University of Arizona, 933 N. 
     Cherry Ave. Tucson, AZ 85721, USA
         \and
 LATT, Universit\'e de Toulouse, CNRS, 14, Avenue Edouard Belin, 31400 Toulouse, France
         \and
 Departamento de Astrof{\'i}sica,  Centro de Astrobiolog{\'i}a, CSIC-INTA, Ctra. Torrej\'on a Ajalvir km 4, 28850 Torrej\'n de Ardoz,
  Spain
         \and
 Max-Planck-Institut f\"ur Astronomie, K\"onigstuhl 17, D-69117 Heidelberg, Germany
         \and
%W.M. Keck Observatory (WMKO), CARA, 65-1120 Mamalahoa Hwy., Kamuela, HI-96743, USA
%         \and
 Institut de Radio Astronomie Millimetrique, Domaine Universitaire, 
    38406 St. Martin d'Heres, France
         \and
 National Radio Astronomy Observatory,
       PO Box 0, Socorro, NM 87801, USA
         \and
 Department of Astronomy, University of Maryland, College Park, 
    MD 20742-2421, USA
         \and
 IRAM, Avenida Divina Pastora, 7, N\'ucleo Central, 
      E-18012 Granada, Spain
         \and
 IPAC, California Institute of Technology, 770 South Wilson Avenue, Pasadena, CA 91125, USA 
             }

\date{Received  / Accepted }

\abstract{We report on a successful, simultaneous observation and modelling 
of the millimeter (mm) to near-infrared (NIR) flare emission of the Sgr$\,$A* counterpart 
associated with the supermassive (4$\times$10$^6$$\,$\solm) black hole at the Galactic centre (GC).  
We present 
a mm/sub-mm light curve of Sgr$\,$A* with one of the highest quality continuous time coverages.}
{We study and model the physical processes giving rise to the variable emission of Sgr$\,$A*.
}
{Our non-relativistic modelling is based on simultaneous observations 
carried out in May 2007 and 2008, using the NACO adaptive
optics (AO) instrument at the ESO's VLT and the mm telescope arrays CARMA in California, ATCA in Australia, and the 30$\,$m IRAM telescope in Spain. We emphasize the importance of multi-wavelength simultaneous fitting as a tool 
for imposing adequate constraints on the flare modelling.
We present a new method for obtaining concatenated light curves of the compact mm-source Sgr$\,$A*
from single dish telescopes and interferometers in the presence of significant 
flux density contributions from an extended and only partially resolved source.}
{The observations detect flaring activity in both the mm domain 
and the NIR.
Inspection and modelling of the light curves  show that in the case of the flare event on 
17 May 2007, the 
mm emission follows the NIR flare emission with a delay of 1.5$\pm$0.5 hours.
On 15 May 2007, the NIR flare emission is also followed by elevated mm-emission.
We explain the flare emission delay
by an adiabatic expansion of source components.
For two other NIR flares, we can only provide an upper limit to any accompanying mm-emission 
of about 0.2$\,$Jy.
The derived physical quantities that describe the flare emission give a source component 
expansion speed of $v_{\mathrm{exp}}$ $\sim$ 0.005$c$~-~0.017$c$, source sizes of about one Schwarzschild radius,
flux densities of a few Janskys, and spectral indices of $\alpha$=0.6 to 1.3. 
These source components peak in the THz regime.
}
{These parameters suggest that either the adiabatically expanding source components
have a bulk motion greater than $v_{\mathrm{exp}}$ or the expanding material 
contributes to a corona or disk, confined to the immediate surroundings of Sgr$\,$A*. 
Applying the flux density values or limits in the mm- and X-ray domain to the observed flare 
events constrains the turnover frequency of the synchrotron components that are on average not lower 
than about 1$\,$THz, such that the optically thick peak flux densities at or below these turnover frequencies do not 
exceed, on average, about $\sim$1$\,$Jy. 
}


\maketitle

\section{Introduction}

Sgr$\,$A*, the compact non-thermal radio and infrared source at the centre of the 
Milky Way galaxy ($\sim$8$\,$kpc away) is known to be associated with a supermassive 
black hole (SMBH) of mass $\sim$4$\times$10$^{6}$$\,$\solm \citep{eckart1996,genzel1997,genzel2000,ghez1998,ghez2000,ghez2004a,ghez2004b,ghez2005,eckart2002,schoedel2002,schoedel2003,eisenhauer2003,eisenhauer2005,gillessen2009,ghez2009}. The close proximity 
of Sgr$\,$A* makes it ideal for studying the evolution and physics of SMBHs 
located in the nuclei of galaxies. The SMBH radiates far below its Eddington 
luminosity at all wavelengths, partly because of its low observed accretion rate. 
For Sgr$\,$A*, we assume that $R_s$=2$R_g$=2$GM$/$c$$^2$$\sim$9$\,$$\mu$as, $R_s$ being one
Schwarzschild radius and $R_g$ the gravitational radius of the SMBH. \\
\indent 
Evidence of flaring activity occurring from a few hours to days has
been found from variability studies ranging from the radio to sub-mm 
wavelengths \citep{bower2002,herrnstein2004,zhao2003,zhao2004,mauerhan2005}. 
There is also evidence that variations in radio/sub-mm emission are 
linked to NIR/X-ray flares, with the radio/sub-mm flares occurring after 
a delay of $\sim$100 minutes after the NIR/X-ray flares 
\citep{eckart2008submm,eckart2006mm,eckart2004,marrone2008,zadeh2008}. 

These flares have been explained with a synchrotron self Compton (SSC)
model that involves up-scattered sub-mm photons from a compact (inferred from short 
flare timescales) source component (e.g., \citealt{eckart2004}, \citealt{eckart2006mm}). 
The X-ray emission is caused by the inverse Compton scattering of the THz-peaked 
flare spectrum by relativistic electrons. Both synchrotron and SSC mechanisms 
contribute to NIR flux density. Adiabatic expansion of the source components 
then gives rise to flares in the mm/sub-mm regimes \citep{eckart2006mm,zadeh2006a,zadeh2006b,zadeh2008,marrone2008}.

Based on the assigned VLT time, we organized an extensive multi-frequency
campaign in May 2007 and May 2008 which included millimeter to NIR observations at
single telescopes and interferometers around the world.  
We present data from observations of Sgr$\,$A* using CARMA\footnote{Support for CARMA 
construction was derived from the states of California, Illinois, and Maryland, 
the Gordon and Betty Moore Foundation, the Kenneth T. and Eileen L. Norris Foundation,
  the Associates of the California Institute of Technology, and the National 
  Science Foundation.  Ongoing CARMA development and operations are supported by
  the National Science Foundation under a cooperative agreement, and by the 
  CARMA partner universities.}\citep{carma2006}, ATCA\footnote{ATCA is operated by the Australia 
  Telescope National Facility, a division of CSIRO, which also includes the ATNF
  Headquarters at Marsfield in Sydney, the Parkes Observatory and the
  Mopra Observatory near Coonabarabran.},  and the MAMBO bolometer at
the IRAM\footnote{The IRAM 30$\,$m millimeter telescope is operated by the
  Institute for Radioastronomy at millimeter wavelengths - Granada,
  Spain, and Grenoble, France.}~30$\,$m telescope in the mm regime and NIR data from the ESO VLT. We detected simultaneous emission
  in the NIR and mm-regimes using the CARMA and ESO VLT telescopes. 
  The main results are: 2 bright NIR flares (16$\,$mJy and 10$\,$mJy), a $\sim$0.4$\,$Jy$\,$mm flare, and a possible weaker third flare in the NIR and mm in the combined ATCA, CARMA, IRAM$\,$30$\,$m  mm/sub-mm light curve 
from May 2007, and a bright NIR flare in May 2008 
also covered by the CARMA 3$\,$mm observations. We present updated versions of light curves first presented in \cite{kunneriath2008}, a more detailed description of the methods used to obtain them, and physical models to explain the flaring activity. 
The observations
  and data reduction are described in Sect. 2, an outline of the flare modelling 
  in Sect. 3 and a discussion and summary follow in Sects. 4 and 5. 
\section{Observations and data reduction}
Interferometric observations in the mm/sub-mm wavelength 
domain are especially well suited to differentiating the flux density 
contribution of Sgr$\,$A* from the thermal emission of the circumnuclear disk 
(CND, a ring-like structure of gas and dust surrounding the Galactic centre at a distance 
of about 1.5-4$\,$pc (see, e.g., \cite{guesten1987} or \cite{christopher2005}). In May 2007 and 2008, global coordinated multiwavelength observations were
carried out in the NIR and mm regimes to study the variability of Sgr$\,$A*.

\subsection{The mm data}
We observed the GC at 100 and 86$\,$GHz (3 and 3.5$\,$mm
wavelength) with the two mm-arrays CARMA and ATCA, respectively.  In addition,
we observed with the MAMBO~2 bolometer at the IRAM 30$\,$m-telescope at a
wavelength of 1.2$\,$mm. CARMA is located in Cedar Flat, Eastern California, and consists
of 15 antennas (6 $\times$ 10.4$\,$m and 9 $\times$ 6.1$\,$m). The quasar source 3C273 was used for bandpass 
calibration while 1733-130 and Uranus were used for phase and amplitude calibration of Sgr$\,$A* 
data, respectively.
The Australia Telescope
Compact Array (ATCA), at the Paul Wild Observatory, is an array of six
22-m telescopes located in Australia. Calibrator sources 1253-055, 1921-293, and Uranus were used 
for bandpass, phase, and amplitude calibration. The
interferometer data were mapped using the {\it Miriad} interferometric data
reduction package.
Details of the observation are given in Table~\ref{log1}.

The Max-Planck Millimeter Bolometer (MAMBO~2) array is installed at
the IRAM~30$\,$m telescope on Pico Veleta, Spain. The 37 channel array
of the precursor instrument MAMBO
has been successfully used by many observers since the end of
1998. The bolometer data was reduced using the bolometer array data
reduction, analysis, and handling software package, the BoA (Bolometer
Data Analysis).

\begin{table*}
\centering
{\begin{small}
\begin{tabular}{cccll}
\hline
Telescope & Instrument/Array & $\lambda$ & UT and JD & UT and JD \\
Observing ID & & & Start Time & Stop Time \\
\hline
ATCA           & H214      & 3.5~mm      & 2007 15 May 07:36:05 & 15 May 22:57:02.5\\
	       &           &             & JD 2454235.81672      & JD 2454236.45628 \\
IRAM~30$\,$m       & MAMBO bolometer&1.2~mm    & 2007 16 May 00:20:42 & 16 May 04:18:56 \\
	       &           &             & JD 2454236.51437      & JD 2454236.67981 \\
CARMA	       & D array   & 3.0~mm	 & 2007 16 May 07:43:31.3 & 16 May 13:27:07.8\\
               &           &             & JD 2454236.82189      & JD 2454237.06051 \\
ATCA           & H214      & 3.5~mm      & 2007 16 May 09:31:57.5 & 16 May 22:21:22.5\\
               &           &             & JD 2454236.89719      & JD 2454237.43151 \\
IRAM~30$\,$m       & MAMBO bolometer&1.2~mm    & 2007 17 May 00:14:39 & 17 May 04:28:13\\
               &           &             & JD 2454237.51017      & JD 2454237.68626 \\	  
CARMA          & D array   & 3.0~mm      & 2007 17 May 07:22:46 & 17 May 13:20:43  \\
               &           &             & JD 2454237.80748      & JD 2454238.05605 \\
ATCA           & H214      & 3.5~mm      & 2007 17 May 09:47:17.5 & 17 May 18:22:32.5\\
               &           &             & JD 2454237.90784      & JD 2454238.26565 \\
IRAM~30$\,$m       & MAMBO bolometer&1.2~mm    & 2007 18 May 00:12:57 & 18 May 04:23:03 \\
               &           &             & JD 2454238.50899      & JD 2454238.68267 \\	       
CARMA	       & D array   & 3.0~mm	 & 2007 18 May 07:33:21.8 & 18 May 13:19:20.3\\
               &           &             & JD 2454238.81484      & JD 2454239.05510 \\
ATCA           & H214      & 3.5~mm      & 2007 18 May 10:11:27.5 & 18 May 22:23:47.5\\
               &           &             & JD 2454238.92462      & JD 2454236.67981 \\	       
CARMA	       & D array   & 3.0~mm	 & 2007 19 May 07:30:52.3 & 19 May 13:15:42.8\\
               &           &             & JD 2454239.81311      & JD 2454240.05258 \\
ATCA           & H214      & 3.5~mm      & 2007 19 May 09:30:57.5 & 19 May 22:13:07.5 \\
               &           &             & JD 2454239.89650      & JD 2454240.42578 \\	       
CARMA          & C array   & 3.0~mm      & 2008 26 May 06:45:13 & 26 May 13:10:43  \\
               &           &             & JD 2454612.78140      & JD 2454613.04911 \\
\hline
\end{tabular}
\end{small}}
\caption{Log of the mm- and sub-mm observations.}
\label{log1}
\end{table*}

\begin{table*}
\centering
{\begin{small}
\begin{tabular}{cccll}
\hline
Telescope & Instrument/Array & $\lambda$ & UT and JD & UT and JD \\
Observing ID & & & Start Time & Stop Time \\
\hline
VLT UT~4       & NACO      & 2.2~$\mu$m  & 2007 15 May 05:29:00 & 15 May 09:42:00 \\
               &           &             & JD 2454235.72847     & JD 2454235.90417 \\
VLT UT~4       & NACO      & 2.2~$\mu$m  & 2007 16 May 04:47:22 & 16 May 07:54:41 \\
               &           &             & JD 2454236.69956     & JD 2454236.82339 \\
VLT UT~4       & NACO      & 2.2~$\mu$m  & 2007 17 May 04:42:14 & 17 May 09:34:40  \\
               &           &             & JD 2454235.69600     & JD 2454235.89907 \\
VLT UT~4       & NACO      & 2.2~$\mu$m  & 2007 19 May 04:55:00 & 19 May 09:28:22 \\
               &           &             & JD 2454239.70486     & JD 2454239.89470 \\
VLT UT~4       & NACO      & 3.8~$\mu$m  & 2007 15 May 10:05:48 & 15 May 10:32:40 \\
               &           &             & JD 2454235.92069     &  JD 2454235.93935 \\
VLT UT~4       & NACO      & 3.8~$\mu$m  & 2007 16 May 08:28:34 & 16 May 10:44:04 \\
               &           &             & JD 2454236.85317     & JD 2454236.94727 \\
VLT UT~4       & NACO      & 3.8~$\mu$m  & 2007 17 May 10:16:24 & 17 May 10:28:00 \\
               &           &             & JD 2454237.92806     & JD 2454237.93611 \\
VLT UT~4       & NACO      & 3.8~$\mu$m  & 2007 18 May 06:03:26 & 18 May 10:29:05 \\
               &           &             & JD 2454238.75238     & JD 2454238.93686 \\
VLT UT~4       & NACO      & 3.8~$\mu$m  & 2007 22 May 04:57:04 & 22 May 06:22:01 \\
               &           &             & JD 2454242.70630     &  JD 2454242.76529  \\
VLT UT~4       & NACO      & 3.8~$\mu$m  & 2007 23 May 04:39:08 & 23 May 10:36:24 \\
               &           &             & JD 2454243.69384     &  JD 2454243.94194  \\ 
VLT UT~4       & NACO      & 3.8~$\mu$m  & 2008 26 May 05:42:55 & 26 May 10:37:39  \\
               &           &             & JD 2454612.73814     & JD 2454612.94281 \\	       
\hline
\end{tabular}
\end{small}}
\caption{Log of the near-infrared observations.}
\label{log2}
\end{table*}
To correct for extended flux contributions in the
interferometer data, we extracted visibilities from two orthogonal pairs of the longest baselines for 
the two arrays, and subtracted for each baseline the median baseline and time dependent visibility trend 

\begin{equation}
D_b(t) = \mu_{\mathrm{epoch}}(d_b(t))
\end{equation}
from each visibility data set $d_b(t)$.
Here the operator $\mu_{\mathrm{epoch}}$ represents the median over all epochs.
The time-dependent differential visibilities 
$S(t)$ and their uncertainties $\delta S(t)$ 
were then calculated via

\begin{equation}
S(t) = \mu_b(d_b(t)-D_b(t))
\end{equation}
and
\begin{equation}
\delta S(t) = \mu_b (S(t) - (d_b(t)-D_b(t)))~~~.
\end{equation}

Here the operator $\mu_b$ is the median over different baselines.
Hence, $\delta S(t)$ is the median of the deviation from the median flux $S(t)$.
The visibilities had been calibrated via
intermittent flux reference observations. We attribute the
  residual flux density dips/excesses to variations in the intrinsic
  flux density of Sgr$\,$A*.  The combined light curve from all
telescopes is shown in Fig. 1.  
Here we can combine the data from different frequencies based on the 
assumption that the spectral index of Sgr$\,$A* does not change 
significantly during the flux density variations between 86 and 250$\,$GHz.
In this case, the flux density variations are frequency independent.

The influence of flux spectral index variations can be calculated in the following way.
If $S_1$ and $S_2$ are the flux densities at the frequencies $\nu_1$ and $\nu_2$,
then a change from spectral index $\alpha$ to $\alpha + \Delta \alpha$ results
in a new flux density value $S'_1$ and with

\begin{equation}
log~S_1 = log~S_2 + \alpha~(log~\nu_1 - log~\nu_2)
\end{equation}
\begin{equation}
log~S'_1 = log~S_2 + (\alpha + \Delta \alpha)(log~\nu_1 - log~\nu_2)
\end{equation}
we find 
\begin{equation}
log \frac{S_1}{S'_1} = - \Delta \alpha~log \frac{\nu_1}{\nu_2} 
\end{equation}
\noindent or for $S'_1 =f S_1$ 
\begin{equation}
log~f = \Delta \alpha~log \frac{\nu_1}{\nu_2}~~. 
\end{equation}
\begin{figure*}[!Htbp]
\begin{center}
\includegraphics[width=18cm, angle=-0]{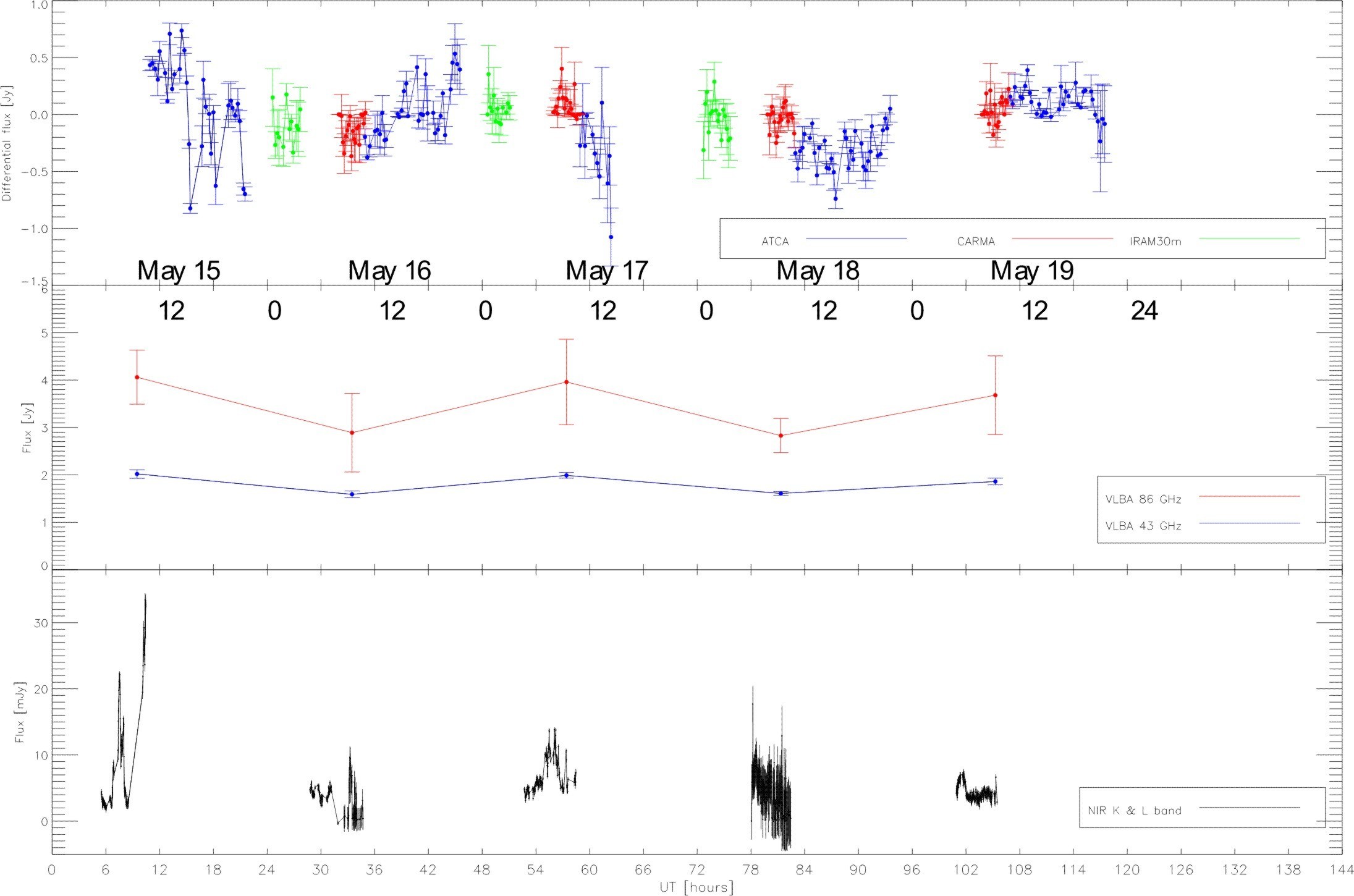}
\end{center}
\caption{ Combined differential light curve of Sgr$\,$A* in the mm/sub-mm
  domain for the May 2007 observing run.  The MAMBO~2 bolometer  at
  the IRAM 30$\,$m-telescope was
  operated at a central wavelength of 1.2$\,$mm (250~GHz).
  The CARMA data were
  centred on 100$\,$GHz and the ATCA data on 86$\,$GHz.  We also show the
  daily 7$\,$mm and 3$\,$mm flux density averages from simultaneous VLBA observations \citep{lu2008,lu2009}. 
In the lower part of the plot, we show the combined NIR light curve consisting 
of K-band and L'-band light curves (see Appendix for individual light curves). The time axis is labelled with UT hours starting at 00$\,$h on May 15.
}
\label{fig:1}
\end{figure*}
We estimate a spectral index variation of $\Delta \alpha = 0.07$ 
between 10$^{10}$$\,$Hz and 10$^{12}$$\,$Hz from the compilation of 
radio to sub-mm measurements by \cite{markoff2001} and \cite{zhao2003}.
The maximum spectral range that we cover with CARMA, ATCA, and the IRAM~30$\,$m is
86$\,$GHz to 250$\,$GHz. In total, this gives a maximum expected flux density variation of 8\%
in the light curve due to spectral index variations between the ATCA, CARMA, and the IRAM~30$\,$m data.
The bulk of the data in which we detected the main flux density excursions is covered by 
the 100$\,$GHz CARMA and the 86$\,$GHz ATCA data. The CARMA and ATCA data follow the VLT data set, 
while the 30$\,$m data do not detect significant flux density excursions preceding the VLT data. 
Here the maximum expected flux density variation
in the light curve due to spectral index variations can only be 0.005$\,$Jy,
which is smaller than $\delta S(t)$.
For the CARMA and IRAM$\,$30$\,$m data, the median flux density variation for
each epoch (day) is $\delta S(t) \sim 0.1$$\,$Jy, and for the ATCA data we find that 
$\delta S(t) \sim 0.2$$\,$Jy. These values represent the approximate range in which the 
datasets can be freely shifted in flux density.\\
\indent The light curve in Fig. 1 contains two peaks, on
May 15 and 17 (there is a weaker, third possible peak on May 19).  
In Fig. 1, we also show the daily flux density averages of the 
7$\,$mm VLBA observations that were conducted in
parallel \citep{lu2008,lu2009,kunneriath2008}. The VLBA data follow the overall trend of the combined
CARMA/ATCA/30$\,$m light-curve very well. The NIR coverage in the K and L'-band is given in the lower 
panel of Fig. \ref{fig:1} (see Appendix for a description of the scaling and the individual light curves for each day).

\begin{figure*}
\centering
\includegraphics[width=18cm, angle=-0]{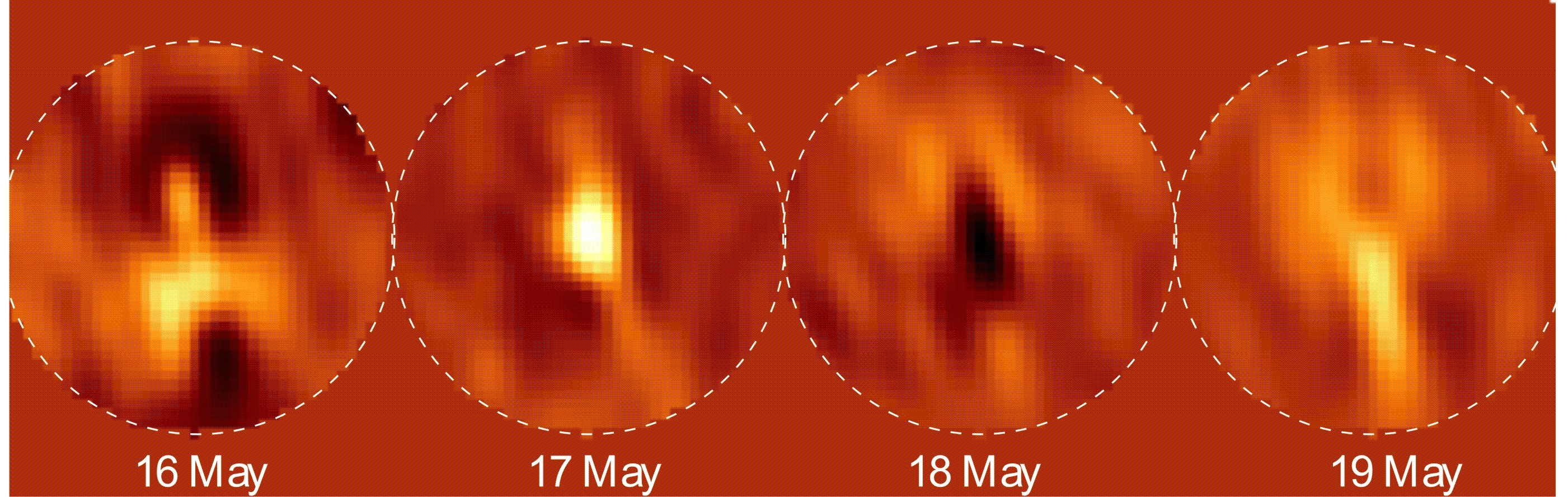}
\caption{Difference maps at 3$\,$mm of a 40$''$ diameter region 
(shown by the dashed white line) centred 
on Sgr$\,$A*, obtained from the difference between full synthesis maps of the
individual days of CARMA observations and the full CARMA data set as described in the text. 
%xxxxxx
The maps have been plotted with the same color-coding table.
The figure shows that the flux density variations that are evident from
the differential light curves (see Fig. \ref{fig:1}) can also be seen in the maps constructed from the
corresponding data.
}
\label{fig:3}
\end{figure*}

To verify the mm/sub-mm flux density variations and in particular 
the flare detected on 17 May by CARMA, we produced residual maps from 
the four individual tracks obtained with the CARMA array, since CARMA has the best combination of high resolution and signal-to-noise ratio of all the mm-telescopes we use. 
These maps shown in Fig. \ref{fig:3}  
were computed by subtracting the mean of all 4 maps from the maps of individual epochs. The rms noise in these differential maps is of the
order of 0.1$\,$Jy per beam. This procedure clearly
represents the trend shown in the combined light curve and 
detects the excess flux density of 0.4$\,$Jy detected on May 17 and a slightly positive flux density on May 19. On the 16th and 18th May 2007, a mixture of negative and positive flux densities in the difference maps corresponds to the non detection of flaring activity in the differential light curve shown in Fig. \ref{fig:1}.

\vspace{2mm}
\noindent

\begin{figure}[!Htbp]
\begin{center}
\includegraphics[scale=0.5,angle=-0,width=9cm]{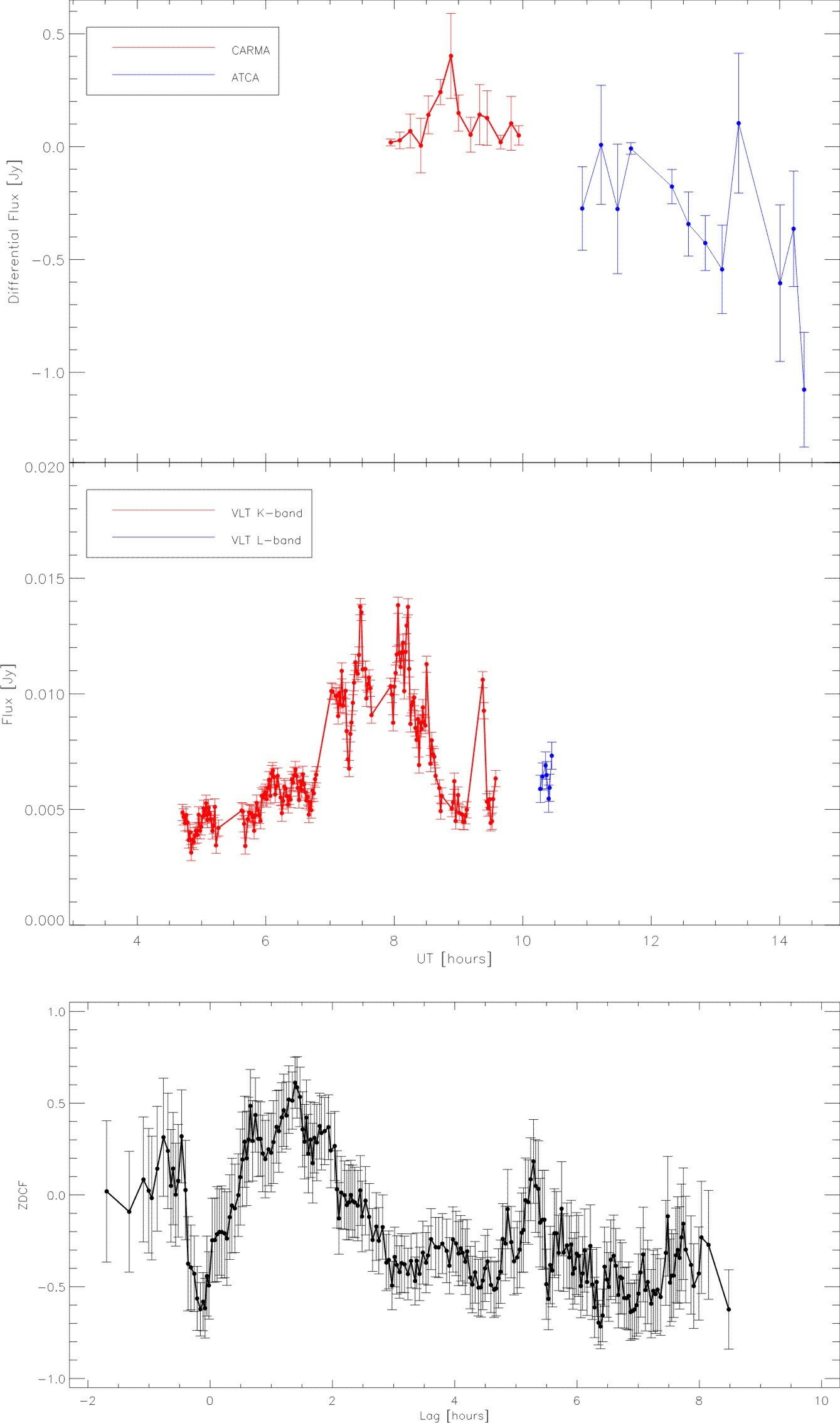}
\caption{
The 3$\,$mm CARMA and ATCA (top red and blue, respectively) and K- and L'-band NIR (middle red and blue, respectively) light curves 
of the 17 May 2007 flare, with the cross-correlation (bottom) between the two showing 
that the mm flare follows the NIR flare with a time lag of 1.5$\pm$0.5 hours. 
ZDCF stands for Z-transformed discrete correlation function.}
\label{flux_obs}
\end{center}
\end{figure}

\begin{figure}[!Htbp]
\begin{center}
\includegraphics[scale=0.5,angle=-0,width=9cm]{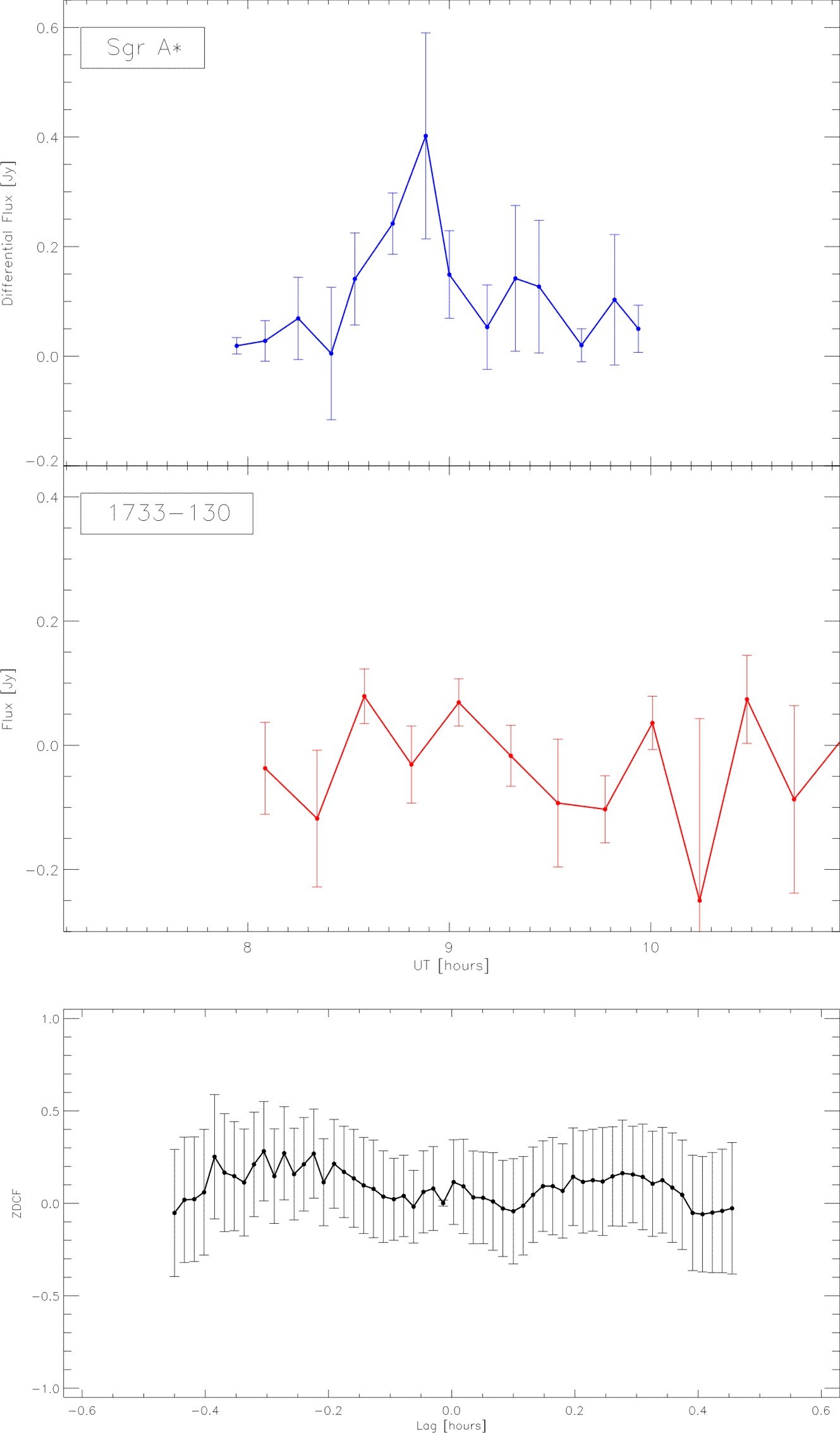}
\caption{3$\,$mm CARMA Sgr$\,$A*(top) and calibrator (1733-130) (middle) light curves from
 17 May 2007, with the cross-correlation (bottom) between the two showing zero
 correlation between the calibrator and the source fluxes. }
\label{flux_calib}
\end{center}
\end{figure}

\subsection{The NIR data}
The NIR data were taken in the K$_S$-band (2.18$\,$$\mu$m, FWHM 0.35$\,$$\mu$m)
and in the L'-band (3.8$\,$$\mu$m, FWHM 0.62$\,$$\mu$m) at ESO's Very Large Telescope (VLT) with the 
NACO infrared camera and adaptive optics (AO) system \citep{rousset2003,lenzen2003} in Chile. 
For the 3.8$\,$$\mu$m observations, the integration times were NDIT$\times$DIT=200$\times$0.2=40 seconds (DIT is Detector 
Integration Time, and NDIT is the number of DIT´s).
At 2.2$\,$$\mu$m, the integration times were NDIT$\times$DIT=4$\times$10=40 seconds.
Additional details of the observation are given in Table~\ref{log2}.
The diffraction limit in K-band is about 60$\,$mas.
The infrared wavefront sensor of NAOS was used to lock
the AO loop on the NIR bright (K-band magnitude $\sim$6.5) supergiant
IRS~7, located about $5.6''$ north of Sgr$\,$A*.  
Therefore the AO  was able to provide a stable correction with a high 
Strehl ratio (of the order of 50\%).
After correcting the images for bad pixels, and subtracting the sky and flat 
fielding, the point spread function (PSF) was extracted from each individual image 
using $\it{StarFinder}$ \citep{diolaiti2000} and 
a Lucy-Richardson (LR) deconvolution was applied. A Gaussian 
beam with FWHM corresponding to the respective wavelength ($\sim$60$\,$mas at 
2.2$\,$$\mu$m and $\sim$104$\,$mas at 3.8$\,$$\mu$m) was used to restore the beam.
To obtain a flux density in Jansky, we perform 
aperture photometry of the sources with circular apertures of radius 52 mas, 
apply extinction corrections of A$_{K}$=2.8 and A$_{L'}$=1.8, and calibrate 
the flux with flux densities of known sources, which resulted in K- and L'-band 
flux densities of S2 of 22$\pm$1$\,$mJy and 9$\pm$1$\,$mJy, respectively. The measurement 
uncertainties were obtained from the reference star S2.   

\subsection{Flare events}
\subsubsection{The 15 May 2007 flare event}
\label{subsubsection:15may2007}
The first NIR flare on May 15 detected during the multiwavelength campaign (see
also \citealt{eckart2008nirpol}) preceded the combined mm/sub-mm monitoring
and the first maximum detected therein by nearly 3 hours. The May 15 2007 flare shows 2 polarized bright sub-flares centred on about
07:29 UT and 07:51 UT, respectively, after the start of observations (see Fig. \ref{fig:1x} in Appendix).
We define sub-flares to be the shorter
flux excursions superimposed on the longer main underlying flare. 
The time difference between the two sub-flares is 22
minutes which is fully consistent with previously reported sub-flare separations. 
\cite{zamani2010} demonstrated that these highly
polarized sub-flares are part of the flare structure
that is significant compared to the randomly polarized red-noise.

The second sub-flare shows substructure that is interpreted by 
\cite{eckart2008nirpol} in the framework of spot evolution.
The flux density between the 2 sub-flares does not
reach the emission level well before and after (i.e., $<$70 and $>$170
minutes into the observations) the flare, probably because of 
fore- or background stellar flux density contributions (star S17).
Additional details of the May 15 2007 NIR flare were given by \cite{eckart2008nirpol}.
%xxxxxxxxx

For the flux density limit reached in low luminosity states of Sgr$\,$A*,
the reader is referred to \cite{sabha2010} and \cite{do2009}.
The brief L'-band observations at 10~UT show a NIR signal that is a factor of
2 times stronger than at other times
(see Sect. \ref{subsubsection:adiab-modelling15May2007}).

\subsubsection{The 17 May 2007 flare event}
\label{subsubsection:17may2007}
The NIR flare on 17 May had overlap with the CARMA observations, 
which indicate that a 0.4$\,$Jy 3$\,$mm flare followed the NIR flare. 

Figure \ref{flux_obs} gives the updated differential light curve
obtained from the four longest baselines of CARMA and ATCA, along with the NIR light curve and 
a cross-correlation between the two showing a time lag of 1.5$\pm$0.5 hours, with 
a significant, broad and positive peak corresponding to the flares found at both wavelengths.
The presence of negative power in the cross-correlation is most likely due to a combination
of two effects: (1) The signal-to-noise ratio of the differential ATCA data is lower than that of the CARMA data;
and (2) the negative power may also be indicative of
residual power on timescales of the length of the (sub-)mm-datasets.
Some amount of negative power is expected since we are dealing with differential 
light curves that measure the variable flux with respect to a median flux level over the
length of the corresponding dataset.
A significant, broad, and positive cross-correlation peak centred on a time lag of $\sim$1.5 hours 
is also present if the correlation is only performed with the K-band and CARMA data.
Figure \ref{flux_calib} shows 
that the calibrator and source fluxes are not correlated for the CARMA data.
Here the cross-correlation was carried out only for the CARMA data since the 
ATCA array did not observe the same calibrator interleaved with the Sgr$\,$A* data.

The NIR K-band data show variable emission, 
with four sub-flares starting from about 
7:00 UT and lasting until the end of the observations at 9:12 UT. 
The mm-data started at about 8:00 UT and had ended by 10:00 UT.  
In the mm domain, the variable emission is dominated by a single 
flare that is slightly wider than the individual NIR sub-flares. 

\subsubsection{The 19 May 2007 flare event}
\label{subsubsection:19may2007}
The flux density excursions on 19 May are only partially constrained by the
CARMA millimeter measurements.
Two flare events of about 3.5$\,$mJy and 1.1$\,$mJy peak flux density are separated by 
about 3 hours. The gap in the observations at a time during which the first event 
reached its peak flux density (see Appendix for individual light curve) implies that the true peak was missed and that the
flare was probably brighter than 3.5$\,$mJy.
There is an overlap between the NIR and the mm-measurements of about 1.5 hours
with no detection of a flare event in the mm-wavelength domain 
brighter than 0.1-0.2$\,$Jy.

\subsubsection{The 26 May 2008 flare event}
\label{subsubsection:may2008}
\begin{figure}[!Htbp]
\begin{center}
\includegraphics[width=8cm]{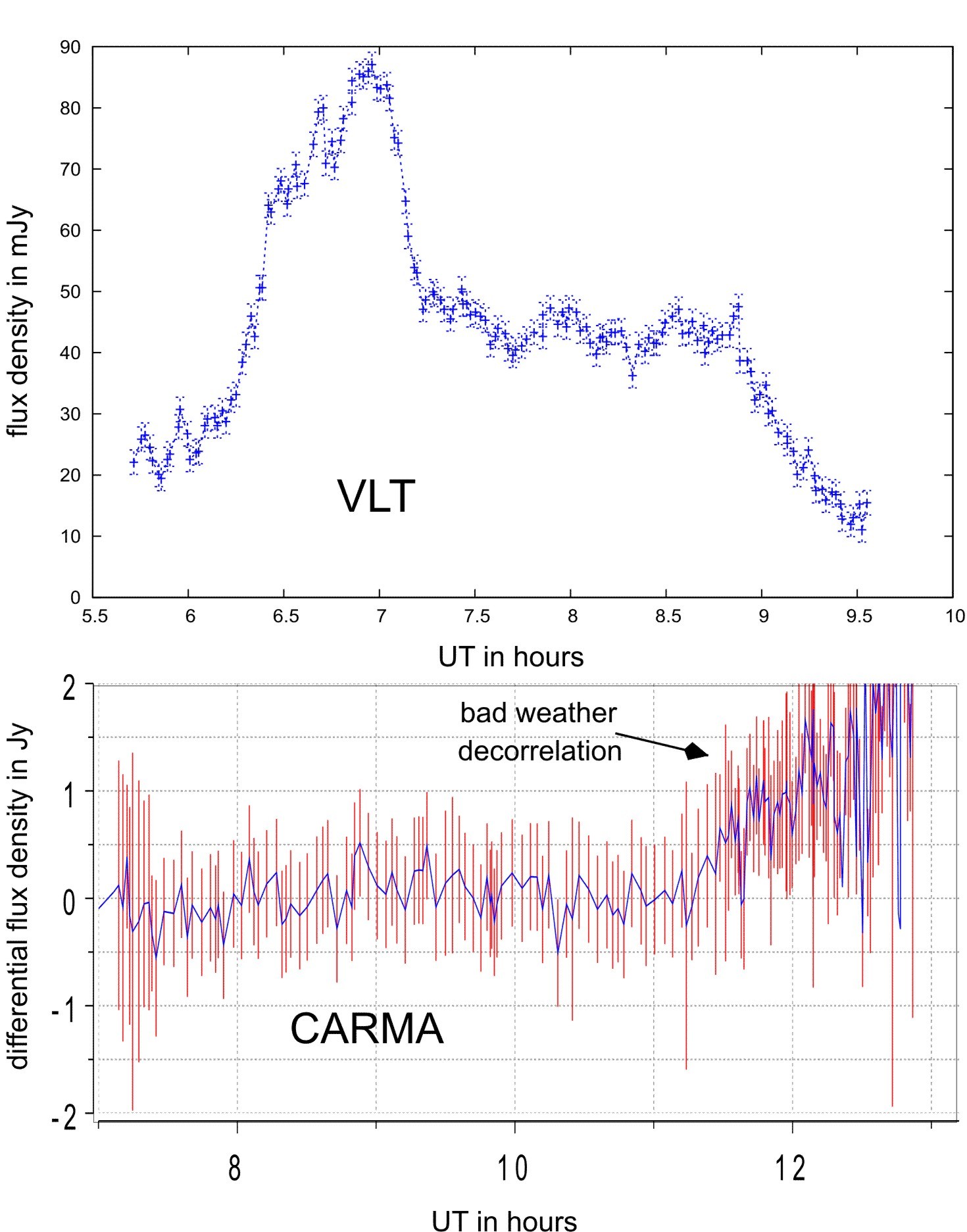}
\caption{3.8$\,$$\mu$m NIR flare (top) covered by 3$\,$mm CARMA data (bottom) on the 26 May, 2008.
The time coverage of the L'-band flare is from 05:42:55 to 10:37:39 UT.}
\label{2008flare}
\end{center}
\end{figure}

The 26 May 2008 NIR flare in the L'-band was covered by the CARMA observations, as 
shown in Fig. \ref{2008flare}. The L'-band flare shows a flux density increase of about 70$\,$mJy, followed by
a plateau with almost constant flux density at a 40$\,$mJy level. After a total flare
duration of 200 minutes, the flux density reaches the original level of 20$\,$mJy.
The CARMA data were taken under moderate to poor weather conditions. From 07:10~UT to 11:10~UT, the 
source visibility was comparable to the levels reached in the 2007 observing session. 
No flare was detected with a flux density limit of about 0.2$\,$Jy.
Starting at 11:10~UT, the weather conditions deteriorated resulting in strong coherence losses.

\section{Flare analysis}
\label{section:FlareAnalysis}
The seven coordinated Sgr$\,$A* measurements reported so far 
that include sub-mm/mm data 
(\citealp{eckart2006mm,eckart2008submm,zadeh2006b,marrone2008,zadeh2009},
including this work)
have shown that the observed sub-millimeter/mm flares follow 
the largest event observed at shorter wavelengths
(NIR/X-ray; see detailed discussion in \cite{marrone2008}) with a delay of $\sim$1.5$\pm$0.5 hours.
We therefore assume that the millimeter 
flares presented here are related to the observed IR flare events.
This flaring activity can be explained using 
a multi-component relativistic spot/disk model and an adiabatic 
 expansion model in combination with a synchrotron self Compton (SSC) 
 formalism. While this model has been presented before (see references above),
it is essential to test new datasets against it to improve
our understanding of the variable emission from Sgr$\,$A*.

\subsection{Relativistic spot/disk modelling of NIR flares}
\label{subsection:Relativistic}
Multiple hot spots revolving around a black hole in an accretion disk in AGNs or galactic black holes can give rise to light curves in the X-ray and NIR regimes whose power spectral density (PSD) is described by a broken power law with a slope similar to that of red noise processes \citep{armitage2003,pechacek2008,do2008}. Simulations show that polarized light curves exhibit behaviour associated with lensing of hot spots orbiting around a SMBH \citep{zamani2010}. Previously observed NIR light curves of Sgr$\,$A* have been successfully modelled by a multi-component hot spot model involving source components orbiting around the SMBH in a temporary accretion disk \citep{eckart2006nirpol,meyer2006a,meyer2006b,meyer2007,zamani2008a,zamani2008b}. 

\begin{figure*}
\centering
\includegraphics[width=18cm, angle=-0]{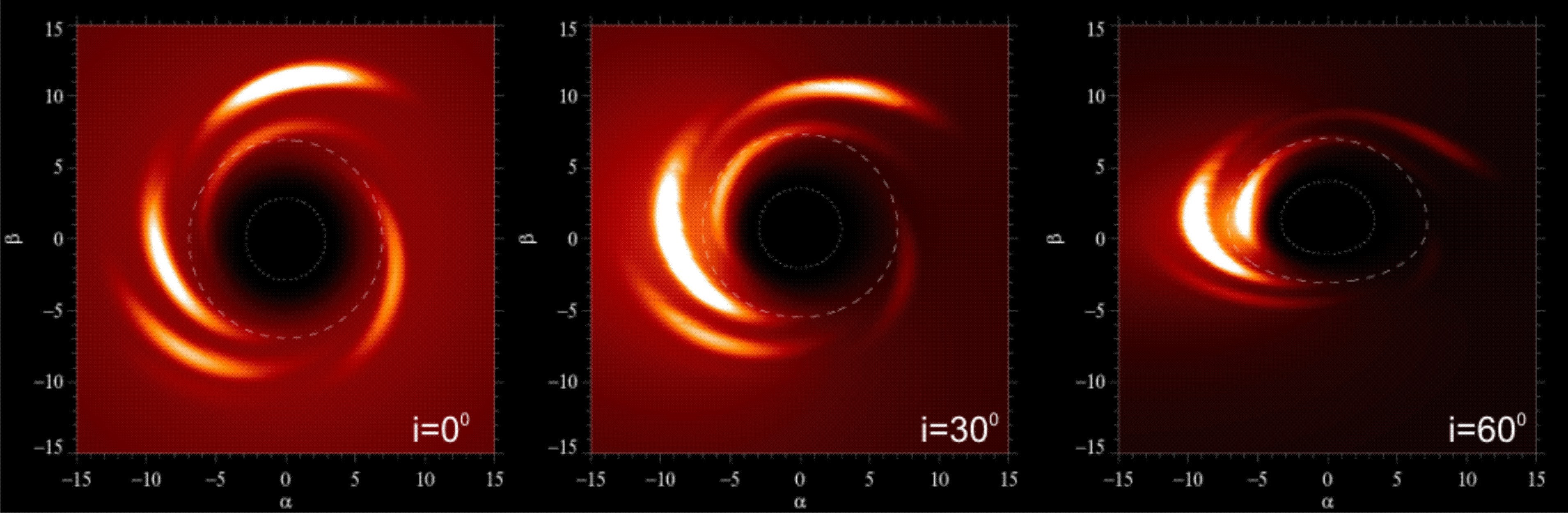}
\caption{Simulated images of multiple spots revolving around a black hole, projected on the observer image plane ($\alpha$, $\beta$), viewed at three inclination angles i=0$^0$, 30$^0$, and 60$^0$, with respect to the common rotation axis. The dotted and dashed circles represent the event horizon and the marginally stable orbit, respectively. 
%xxxxxx
}
\label{fig:hotspot}
\end{figure*}

\begin{figure*}
\centering
\includegraphics[width=19cm, angle=-0]{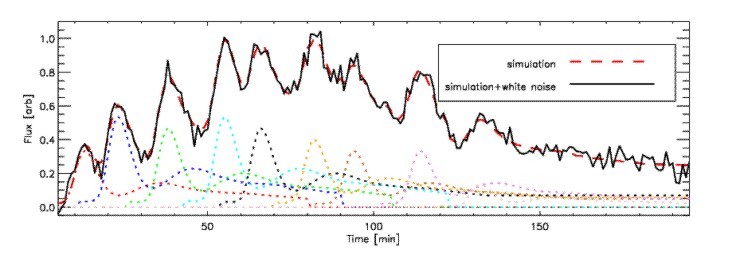}
\caption{Light curve obtained from the multiple spot simulations shown in Fig. \ref{fig:hotspot}, the solid black line representing the overall light curve and the different coloured dashed lines showing the contributions of individual spots. 
%xxxxxx
}
\label{fig:lightcurve}
\end{figure*}

This relativistic model consists of source components that are a mix of synchrotron and SSC components with an optically thin spectral index $\alpha$ and relativistic electrons with $\gamma$$_e$$\sim$10$^3$, 
  using the formalism described by \cite{gould1979} and \cite{marscher1983}. 
The KY-code by \cite{dovciak2004} produces light amplification curves for each individual component 
  orbiting the SMBH taking into account relativistic effects, and by combining these curves with the SSC model, we can 
  estimate the SSC X-ray and NIR flux densities and the magnetic field \citep{eckart2008nirpol}. 
  Figure \ref{fig:hotspot} presents simulations of the accretion disk with multiple spots revolving around the SMBH, and Fig. \ref{fig:lightcurve} gives the light curves produced by these simulations. \cite{zamani2010} describe the modelling and simulations shown in Figs. \ref{fig:hotspot} and \ref{fig:lightcurve} in greater detail. 
%xxxxxxxxxxx

These light curves may be the result of a single hot spot that
dominates the disk for several orbital periods.
If the spot were to sink (as may occur) towards the centre while orbiting, this would cause a quasi-periodic modulation of the light curve for the
time the emission of the spot dominates the red-noise emission of the
remaining disk.
This modulation may also be produced by several different hot spots that have
individual lifetimes of shorter than one orbital period close to the
last stable orbit.
In this case, the light-curve will be modulated quasi-periodically at
a rate close to the frequency of the last stable orbit
(see discussion in \citealt{eckart2008nirpol} and \citealt{eckart2006nirpol}).
If the hot spots expand within the accretion disk or are the source of a short
outflow above the disk,
this quasi-periodic signal will be smeared out. In particular, it
may be undetectable at lower (sub-)mm observing frequencies (see below).
In the case of our experiment, only the NIR data on 15 May shows a strong
sub-flare modulation (as explained below) that may be linked to 
quasi-periodicity. All other NIR flares
(with the possible exception of the steeply rising and falling
flanks of the L'-band flare on 26 May 2008) cannot be associated with
this phenomenon.
However, in the case of a source expansion this model can be linked to the
adiabatic expansion model for synchrotron sources that is described
in the following section.

\vspace{2mm}
\noindent
\vspace{2mm}
\noindent

\subsection{Adiabatically expanding source components}
\label{subsection:Adiabatic}
Adiabatic expansion of the synchrotron components can explain 
the apparent time difference of 1.5$\pm$0.5 hours between the NIR and sub-mm/mm flares \citep{eckart2006mm,eckart2008submm,zadeh2006a}. We assume a uniformly expanding blob of relativistic electrons with a power-law energy spectrum,
 $n(E) \propto E^{-p}$, threaded by a magnetic field that declines as
 $R^{-2}$, and of energy and density that decline as $R^{-1}$ and $R^{-3}$, respectively, as a 
 result of expansion of the blob \citep{laan1966}.
If $R_0$, $S_0$ and $\tau$$_0$ are the size, flux density, and optical depth of the source component at the
peak frequency ($\nu$$_0$) of the synchrotron spectrum, the optical depth and flux density at a given 
frequency $\nu$ scale as 
\begin{equation}
    \tau_\nu = \tau_0 \left(\frac{\nu}{\nu_0}\right)^{-(p+4)/2}
    \left(\frac{R}{R_0}\right)^{-(2p+3)}
    \label{eq:taunu}
\end{equation}
and 
\begin{equation}
    S_\nu = S_0 \left(\frac{\nu}{\nu_0}\right)^{5/2} 
    \left(\frac{R}{R_0}\right)^3 
    \frac{1-\exp(-\tau_\nu)}{1-\exp(-\tau_0)},
    \label{eq:Snu}
\end{equation}
respectively.
Here $\tau$$_0$ is defined to be the optical depth corresponding to 
the frequency at which the flux density is maximum, in accordance with van der Laan's model (1966), to be able to combine the SSC formalism with an adiabatic expansion model.This
 makes $\tau$$_0$ dependent only on $p$ through the condition
 \begin{equation}
    e^{\tau_0} - \tau_0(p + 4)/5 - 1 = 0.
    \label{eq:tau0}
\end{equation}
For instance, if $p$ ranges from 1 to 3, $\tau$$_0$ ranges from 0 to 0.65. 
Thus for a given particle spectral index $p$ and peak flux density $S_0$ at a frequency $\nu$$_0$, 
the model gives the variation in flux density at any frequency as a function of the expansion factor $(R/R_0)$.
To convert the dependence on radius to a dependence on time, we assume a simple linear expansion model with
 a constant expansion speed of $v_{\mathrm{exp}}$, such that $R$-$R$$_0$=v$_{\mathrm{exp}}$($t$-$t_0$).
 At times $t$$<$$t_0$, we assume that the source has an optical depth equal to its frequency-dependent
  initial value $\tau$$_0$ at $R$=$R_0$. For the optically thin part of the spectrum, the flux increases initially with 
  increasing source size at constant optical depth $\tau$$_0$ and then decreases with decreasing optical depth as it expands. 
  One Schwarzschild radius corresponds to $R_s$=2$GM$/$c$$^2$$\sim$10$^{10}$$\,$m for a $\sim$4$\times$10$^6$$\,$\solm supermassive black hole, 
  which infers the velocity of light to be about 100$R_s$ per hour. 
  For $t$$>$$t_0$, increasing the turnover frequency $\nu$$_0$ or initial source size $R_0$ has the effect of shifting the decaying flank 
  of the curve to later times, as does decreasing the spectral index $\alpha_{\mathrm{synch}}$ or peak flux density $S_0$.
  Increasing the adiabatic expansion velocity v$_{\mathrm{exp}}$, on the other hand, shifts the peak of the light curve to earlier times. 
  Flare timescales are longer at lower frequencies and have a slower decay rate, as a result of adiabatic expansion.

%-------------------------------------------------------------
\begin{table*}
\centering
{\begin{small}
\begin{tabular}{lllclrrrrrrrrrr}
\hline
date & model & source   & $\Delta$$t$ &$v_{\mathrm{exp}}$ &$S_{\mathrm{max, obs}}$&$\alpha_{\mathrm{synch}}$&$R_0$  &$\nu_{\mathrm{max, obs}}$ & $B$    &$S_{\mathrm{NIR, synch}}$ &$S_{\mathrm{NIR, SSC}}$  &$S_{\mathrm{X-ray, SSC}}$\\
     & label &          &    hours  & in $c$   &[Jy]          &                &[R$_s$]& [GHz]           & [G]  & [mJy]           &[mJy]           &[nJy]           \\
 \hline
     &       &          &           &          &              &                &       &                 &      &                 &                &           \\
$1~\sigma$ $\rightarrow$  &   &   & $\pm$1.0 &$\pm$0.001 &$\pm$0.1&$\pm$0.1 &$\pm$0.1 &$\pm$250 &$\pm$10&$\pm$1.0&$\pm$1.0 &$\pm$20 \\
     &       &          &           &          &              &                &       &                 &      &                 &                &           \\
 \hline
     &       &          &           &          &              &                &       &                 &      &                 &                &           \\
15 May 2007& A &$\alpha$&   0.0     & 0.007    & 0.6          & 0.85           & 1.3   & 340             & 33   &  3.1            &  $<$0.03       & $<$10  \\
           &   &$\beta$ &   0.0     &          & 0.3          & 0.60           & 0.3   & 840             & 29   &  12             &  $<$0.03       & $<$10  \\
           &   &$\gamma$&   0.30    &          & 0.3          & 0.75           & 0.2   & 840             & 32   &  5.6            &  $<$0.03       & $<$10  \\
           &   &        &           &          &              &                &       &                 &      &                 &                &           \\
15 May 2007& B &$\alpha$&   0.0     & 0.017    & 0.6          & 0.65           & 1.3   & 340             & 37   &  8.3            &  $<$0.02       & $<$40  \\
           &   &$\beta$ &   0.0     &          & 0.3          & 0.65           & 0.3   & 840             & 30   &  8.4            &  $<$0.02       & $<$40  \\
           &   &$\gamma$&   0.37    &          & 0.3          & 0.65           & 0.35  & 810             & 32   &  8.4            &  $<$0.02       & $<$40  \\
           &   &        &           &          &              &                &       &                 &      &                 &                &           \\
17 May 2007& A &$\alpha$&   0.0     & 0.010    & 1.0          & 1.00           & 0.8   &  840            & 66   &  4.7            &  $<$0.01       & $<$10  \\
           &   &$\beta$ &   0.45    &          & 0.5          & 0.97           & 0.3   & 1250            & 66   &  4.0            &  $<$0.01       & 15  \\
           &   &$\gamma$&  -0.59    &          & 1.3          & 1.11           & 1.0   &  840            & 69   &  3.2            &  $<$0.01       & $<$10  \\
           &   &$\delta$&  -0.96    &          & 0.5          & 1.02           & 0.3   & 1250            & 68   &  3.2            &  $<$0.01       & 12  \\
           &   &        &           &          &              &                &       &                 &      &                 &                &           \\
17 May 2007& B &$\alpha$&   0.0     & 0.011    & 0.25         & 0.70           & 0.2   & 1360            & 86   &  8.1            &  $<$0.01       & 19  \\
           &   &$\beta$ &   0.36    &          & 0.25         & 0.70           & 0.2   & 1360            & 86   &  8.1            &  $<$0.01       & 19  \\
           &   &$\gamma$&  -0.69    &          & 0.17         & 0.70           & 0.2   & 1360            & 65   &  5.1            &  $<$0.01       & 21  \\
           &   &$\delta$&  -1.05    &          & 0.17         & 0.70           & 0.2   & 1360            & 65   &  5.1            &  $<$0.01       & 21  \\
           &   &$\epsilon$&-0.33    &          & 1.75         & 1.05           & 1.6   &  570            & 68   &  7.1            &  $<$0.01       & $<$2  \\
           &   &        &           &          &              &                &       &                 &      &                 &                &           \\
19 May 2007& A &$\alpha$&   0.0     & 0.007    & 1.30         & 1.07           & 1.0   & 1360            & 67   &  3.5            &  $<$0.01       & $<$10  \\
           &   &$\beta$ &   2.50    &          & 1.10         & 1.10           & 0.8   & 1360            & 67   &  0.5            &  $<$0.01       & $<$10  \\
           &   &$\gamma$&   2.90    &          & 1.10         & 1.10           & 0.8   & 1360            & 67   &  0.5            &  $<$0.01       & $<$10  \\
           &   &        &           &          &              &                &       &                 &      &                 &                &           \\
19 May 2007& B &$\alpha$&   0.0     & 0.017    & 1.33         & 1.13           & 1.0   &  720            & 68   &  2.5            &  $<$0.01       & $<$10  \\
           &   &$\beta$ &   0.50    &          & 1.33         & 1.09           & 1.0   &  720            & 67   &  3.1            &  $<$0.01       & $<$10  \\
           &   &$\gamma$&   2.85    &          & 1.50         & 1.30           & 0.8   &  820            & 72   &  1.2            &  $<$0.02       & $<$10  \\
           &   &$\delta$ &   3.30    &          & 1.50         & 1.30           & 0.8   &  850            & 72   &  1.2            &  $<$0.02       & $<$10  \\
           &   &        &           &          &              &                &       &                 &      &                 &                &           \\
26 May 2008& A &$\alpha$&   0.0     & 0.005    & 1.1          & 0.68           & 0.5   & 1090            & 30   &  31             &  $<$0.02       & 41  \\
           &   &$\beta$ &  -0.45    &          & 1.0          & 0.68           & 0.5   & 1090            & 40   &  26             &  $<$0.02       & 18  \\
           &   &$\gamma$&   0.52    &          & 1.0          & 0.83           & 0.5   & 1030            & 31   &  15             &  $<$0.02       & 17  \\
           &   &$\delta$&   1.03    &          & 1.0          & 0.80           & 0.5   & 1160            & 34   &  17             &  $<$0.03       & 23  \\
           &   &$\epsilon$& 1.48    &          & 0.8          & 0.77           & 0.5   & 1160            & 48   &  16             &  $<$0.01       & $<$10  \\
           &   &$\zeta$ &   1.90    &          & 0.8          & 0.77           & 0.5   & 1160            & 48   &  16             &  $<$0.01       & $<$10  \\
           &   &        &           &          &              &                &       &                 &      &                 &                &           \\
26 May 2008& B &$\alpha$&   0.0     & 0.007    & 1.4          & 0.70           & 0.6   & 1090            & 44   &  39             &  $<$0.02       & 24  \\
           &   &$\beta$ &  -0.40    &          & 1.3          & 0.70           & 0.6   & 1090            & 56   &  35             &  $<$0.02       & 13  \\
           &   &$\gamma$&   0.52    &          & 1.1          & 0.77           & 0.6   & 1030            & 36   &  23             &  $<$0.02       & 18  \\
           &   &$\delta$&   1.03    &          & 1.3          & 0.80           & 0.5   & 1160            & 33   &  23             &  $<$0.04       & 33  \\
           &   &$\epsilon$& 1.48    &          & 1.0          & 0.77           & 0.5   & 1160            & 58   &  20             &  $<$0.01       & $<$10  \\
           &   &$\zeta$ &   1.90    &          & 1.0          & 0.73           & 0.5   & 1160            & 56   &  24             &  $<$0.01       & $<$10  \\
           &   &        &           &          &              &                &       &                 &      &                 &                &           \\

 \hline 
\end{tabular}
\end{small}}
\caption{
Source component parameters for the combined SSC and adiabatic expansion model
of the 15, 17, and 19 May 2007 and the 26 May 2008 flares. Labels A and B refer to models with lower and higher expansion velocities, respectively. 
The flare times $\Delta$$t$ are given with respect to the peak of the brighter NIR flares.
In addition to $v_{\mathrm{exp}}$, the $R_0$ values are responsible for the 
position and width of the infrared flare peaks in time. 
Different values for $\alpha_{\mathrm{synch}}$ are required to match 
the infrared flux densities.
\label{modeldata}}
\end{table*}
\subsection{Adiabatic expansion modelling}
\label{subsection:adiab}
Table~\ref{modeldata} summarises the properties of the model, such as the adiabatic expansion velocity $v_{\mathrm{exp}}$, the optically thin spectral index $\alpha$$_{\mathrm{synch}}$, cutoff frequency $\nu$$_{\mathrm{max,obs}}$, and flux density $S$$_{\mathrm{max,obs}}$ of the source components, using the same nomenclature as  \cite{eckart2006mm,eckart2009}. The optically thin NIR flux density is represented by $S$$_{\mathrm{NIR,synch}}$, while $S$$_{\mathrm{NIR,SSC}}$ and $S$$_{\mathrm{X-ray,SSC}}$ give the upper limits to the flux densities of upscattered SSC components.
A global variation in a single parameter 
by the value listed in the corresponding column in Table~\ref{modeldata}
 results in an increase of $\Delta \chi = 1$ (from the reduced $\chi^2$ value).
Here global variation  means adding to a single model parameter for all source components the 1$\sigma$
uncertainty, such that a maximum positive or negative flux density deviation is reached.
Alternatively, a variation by the listed uncertainty for 
only a single source component results in a variation of the model
predicted NIR and X-ray flux density by more than 30\%.
Judging from the $\Delta \chi$ based on the  mm-data alone,
the global uncertainties for $S_{\mathrm{max,obs}}$, $\alpha_{\mathrm{synch}}$, and $R_0$ 
may then be doubled.\\
When performing the reduced $\chi^2$ fit for $N$ source components, 
we used $N$ times 4 ($S_{\mathrm{max,obs}}$, $\alpha_{\mathrm{synch}}$,
$R_0$, $\nu_\mathrm{{max}}$) plus one common expansion velocity $v_{\mathrm{exp}}$ and
time offset (leaving the time differences between the components fixed),
i.e., 4$N$+2 degrees of freedom.
Unfortunately, the model parameters may not all be considered 
independent, e.g., the width and peak of a light curve signature
depends to a varying extent on all 4 parameters $S_{\mathrm{max}}$, $\alpha_{\mathrm{synch}}$,
$R_0$, and $\nu_{\mathrm{max}}$. Therefore, for all models we stayed 
with the minimum number of source components to estimate
the degrees of freedom. 
Since in the case of the May 2007 flare the VLT data consist of $\sim$10 times 
the number of CARMA data points,
we weighted the squared CARMA flux deviations and number of data points
by an additional factor of 19. The $\chi^2$ test was then carried out using the
sum of the squared flux deviations and data points of the VLT and CARMA 
datasets.\\
In general, we aimed to develop a model at a lower and a higher expansion 
speed labelled A and B (Table~\ref{modeldata}). In comparison to earlier modelling results 
\citep{eckart2008nirpol,eckart2009,zadeh2008,marrone2008},
these speeds were chosen to be 0.007$c$ and 0.017$c$
for the 15 May and 19 May 2007 flares.
For the 17 May 2007 and the 26 May 2008 data, a violation of the observed flare width 
and the expected range for the magnetic field strength restricted the choice of expansion speeds.
For comparison, we listed the model components in Table~\ref{modeldata}.
The models were developed by minimizing the
number of free parameters (and maximizing the description of significant 
flare features in the observed light curves).

{\it SSC modelling as an additional constraint:}
The SSC model described in Sect. \ref{subsection:Relativistic} allows us to estimate 
the SSC contribution of the X-ray and NIR flux densities, and the magnetic field. Higher SSC NIR emission would 
violate the assumption in our adiabatic expansion model that the dominant source of NIR flux density is synchrotron 
emission of the THz peaked expanding component, as well as require us to model the unknown X-ray flux density. 
Magnetic fields of up to $\sim$70 gauss 
\citep{eckart2006mm,eckart2008submm,zadeh2008,marrone2008} 
have been obtained for previous models via 
$B \sim \theta^4 \nu_m^5 S_m^{-2}$, where 
 $\nu$$_m$ and S$_m$ are the synchrotron turnover frequency 
and flux density, respectively. These results should also 
not be violated by the model. \\
\indent {\it The expansion speed:}
The model yields expansion velocities from 0.005$c$~-~0.017$c$ (see Table~\ref{modeldata}), which agree well with 
previously published results \citep{zadeh2008}.
These velocities are lower than expected for relativistic sound speed in orbital velocity near the SMBH. 
This may be caused by the bulk velocity of the source components being larger than the expansion velocity $v_{\mathrm{exp}}$ or
because the expanding material is confined to the immediate region surrounding Sgr$\,$A* in the 
form of a disk or corona. 
In this case, shearing caused by differential rotation within the accretion disk is 
responsible for the 'expansion' and the observed low 
expansion velocities \citep{eckart2008submm,zamani2008a,pechacek2008}.
Owing to the expansion or the presence of several strong spots in the
disk or corona, the short time modulation often seen in the NIR will also
be smeared out in the (sub-)mm-wavelength domain - implying that it
may not be observable at all at these wavelengths.\\
By comparing other flare events that were detected simultaneously
at X-ray and sub-mm/mm-wavelengths, we imposed the additional constraint 
that the 100 GHz flux density should not be much stronger than about 
0.2$\,$Jy, 1.5$\pm$0.5 hours after the NIR event, and
the X-ray flux density should be limited to a few 10$\,$nJy. The following sections describe the modelling of the individual flares.

\begin{figure}[!Htbp]
\begin{center}
\includegraphics[width=8cm, angle=-0]{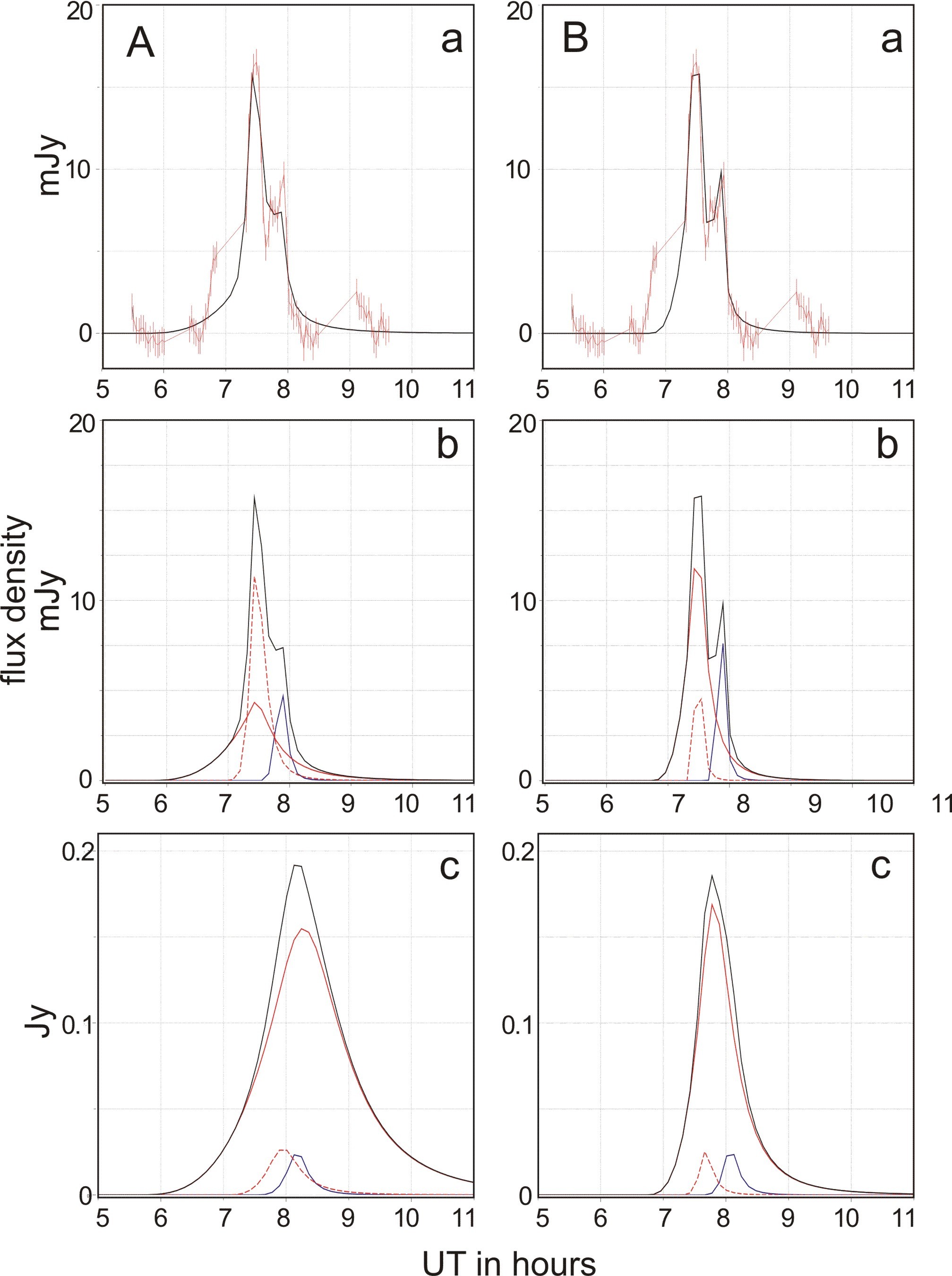}
\end{center}
\caption{Models A and B of the 15 May 2007 NIR flare: a) Fitted NIR light curve, with the black curve corresponding to the model and the red curve corresponding to the data, b) Contribution of individual source components to NIR light curve, and c) Contribution of individual source components to mm light curve.}
\label{fig:model15}
\end{figure}
\begin{figure}[!Htbp]
\begin{center}
\includegraphics[width=8cm, angle=-0]{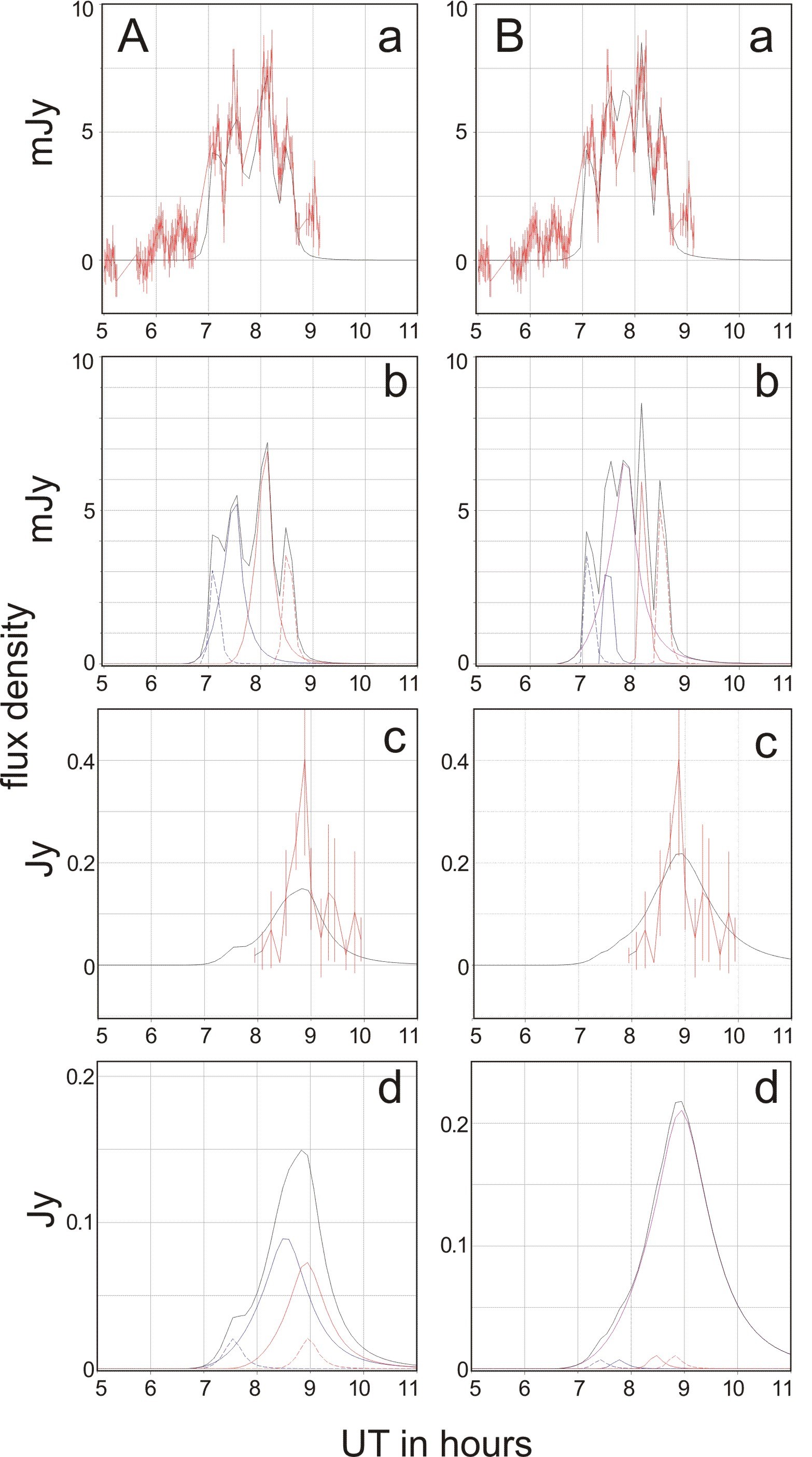}
\end{center}
\caption{Models A and B of the 17 May 2007 NIR flare: a) Fitted NIR light curve, with the black curve corresponding to the model and the red curve corresponding to the data, b) Contribution of individual source components to NIR light curve, c) Fitted mm light curve, with the black curve corresponding to the model and the red curve corresponding to the data, and d) Contribution of individual source components to mm light curve.}
\label{fig:model17}
\end{figure}

\begin{figure}[!Htbp]
\begin{center}
\includegraphics[width=8cm, angle=-0]{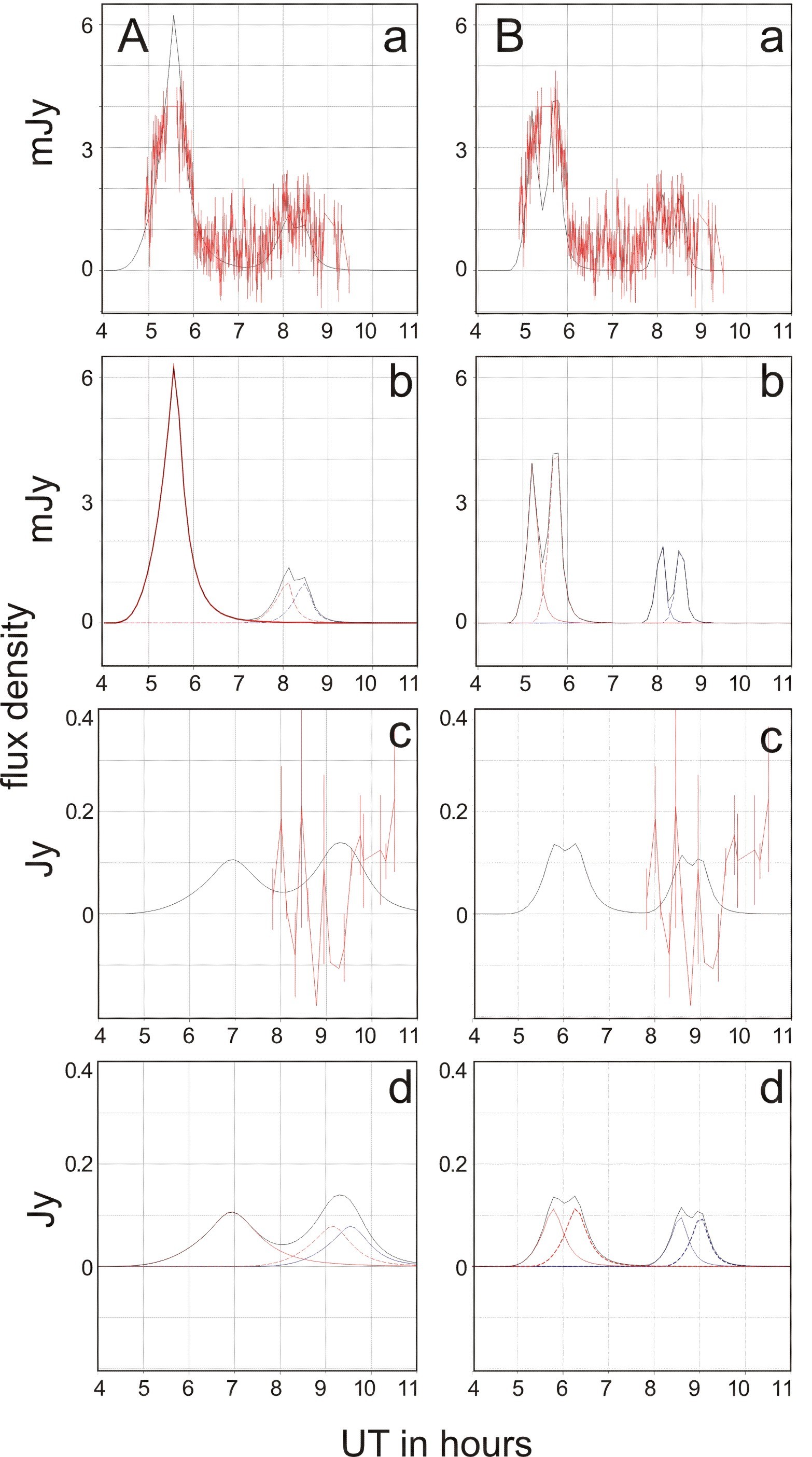}
\end{center}
\caption{Models A and B of the 19 May 2007 NIR flare: a) Fitted NIR light curve, with the black curve corresponding to the model and the red curve corresponding to the data, b) Contribution of individual source components to NIR light curve, c) Fitted mm light curve, with the black curve corresponding to the model and the red curve corresponding to the data, and d) Contribution of individual source components to mm light curve.}
\label{fig:model19}
\end{figure}

\begin{figure}[!Htbp]
\begin{center}
\includegraphics[width=8cm, angle=-0]{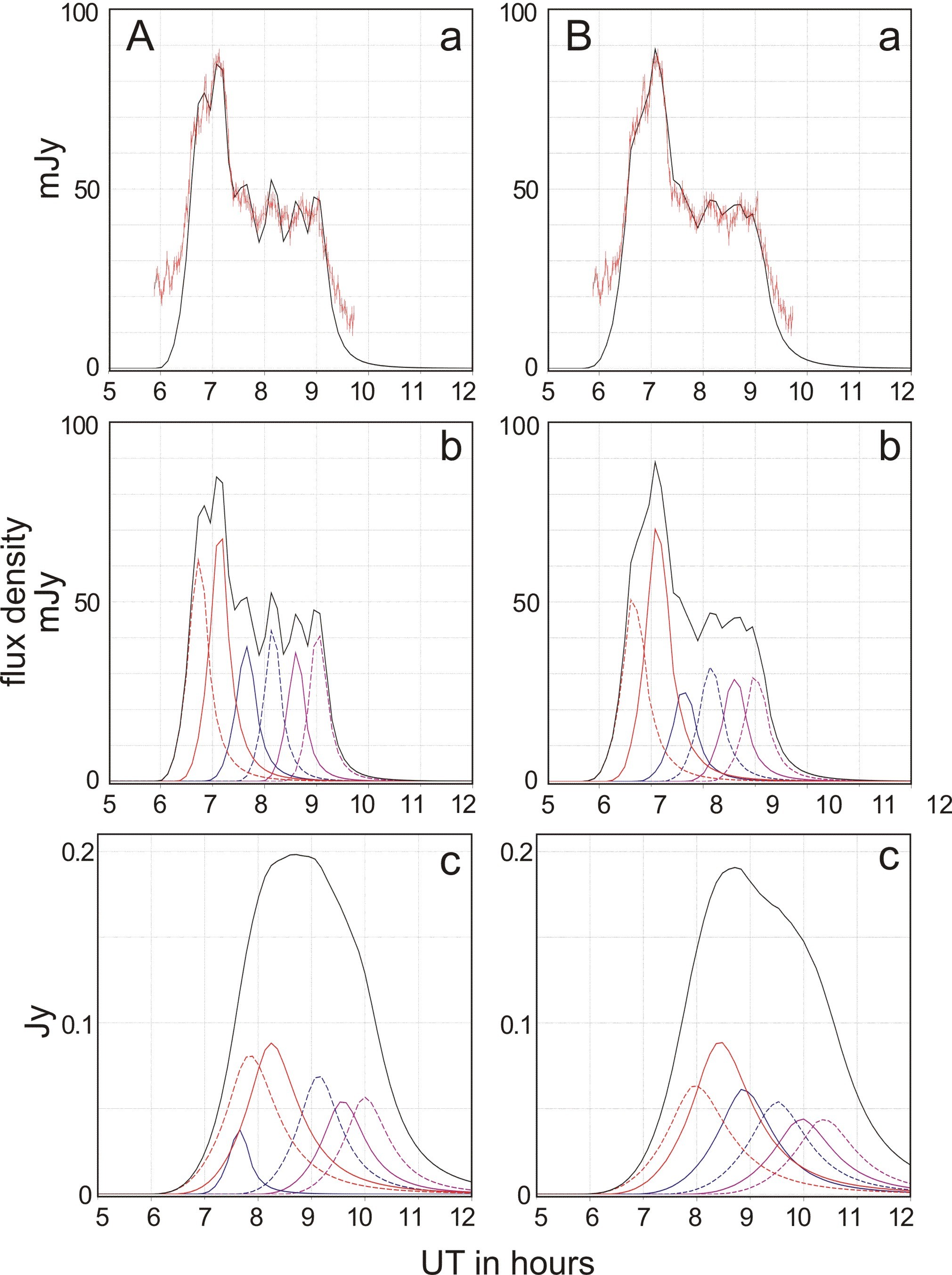}
\end{center}
\caption{Models A and B of the 26 May 2008 NIR flare: a) Fitted NIR light curve, with the black curve corresponding to the model and the  red curve corresponding to the data, b) Contribution of different source components to NIR light curve, and c) Contribution of different source components to mm light curve.}
\label{fig:model26}
\end{figure}

\subsubsection{Expansion modelling of the 15 May flare emission}
\label{subsubsection:adiab-modelling15May2007}
The 15 May 2007 flare can only be constrained by means of its NIR lightcurve 
(Fig. \ref{fig:model15}) since the mm data begins nearly 3 hours after the 
first peak in the NIR. According to the adiabatic expansion model, 3-4 hours after the NIR flare, 
the flux density in the mm band was insufficiently strong to be significant compared to the median flux density variations 
in the data. 
We modelled a background component $\beta$ and two components for the two strongest
sub-flares $\alpha$ and $\gamma$. The expansion speed could not be constrained at all. 
In model A, we used 
a value of 0.007$c$, close to the speed we found for most of the modelled
flares. In model B, we used a higher speed of 0.017$c$, which required lower
spectral indices and a weaker background component to obtain a good fit to the NIR data
and to fullfil the upper limits at 100$\,$GHz and in the X-ray domain.
%xxxxxxxxx

A correlation
between the L'-band data at 10~UT and the radio data cannot be excluded,
but the L'-band data light curve is not long enough to justify including
this data in a
full modelling of the NIR/mm-light curves in the framework of the
adiabatic expansion model. However, high NIR flux density levels
precede a positive flux density excursion in the mm-domain,
as it is expected to in the framework of the adiabatic expansion model.
Although the L'-band flux density is higher than that at other times,
the event is not linked to a particularly strong flux density 
excursion in the mm-domain.
However, the high L'-band flux may also be caused by a transient steeper infrared spectral index. This may indicate either
a significantly higher turnover frequency 
(extrapolating into the sub-mm domain) or the presence of stronger
synchrotron losses. Both effects may explain a lack of strong
mm-emission following the NIR event at 10~UT.

\subsubsection{Expansion modelling of the 17 May flare emission}
\label{subsubsection:adiab-modelling17May2007}
The 17 May 2007 flare is constrained by the NIR and CARMA mm-data (Fig. \ref{fig:model17}). This time lag of 1.5$\pm$0.5 hours between
the flare events mainly determines the expansion velocity. A constant flux density of 0.003$\,$Jy was subtracted from the 
NIR data to model the individual flare events. We present two modelling
approaches that lead to a successful description of the flare. In model A, we used only
4 components describing 4 sub-flares $\alpha$ through $\delta$. Motivated by the 
relativistic disk modelling referred to in Sect. \ref{subsection:Relativistic},
we used in model B a
background component $\epsilon$ located close to the centre of the overall NIR flare.
We then modelled the 4 sub-flares with additional components. In comparison to model A, this
results in lower values for $S_{\mathrm{max,obs}}$ and the spectral index 
$\alpha_{\mathrm{synch}}$, smaller source component sizes $R_0$, and higher cutoff frequencies $\nu_{\mathrm{max,obs}}$ for components $\alpha$ through $\delta$.
The quality of the fit is comparable for models A and B.

\subsubsection{Expansion modelling of the 19 May flare emission}
\label{subsubsection:Orbitmodelling19May2007}
For the 19 May 2007 flare, the slower adiabatic expansion velocity of 0.007$c$ allows us to fit the 
first event (centred on the NIR at $\sim$5:30~UT) with a single component (Fig. \ref{fig:model19}). 
For higher velocities, a larger number of source components has to be used to model the NIR data.
For both velocities, the second, later flare component (centred on the NIR at $\sim$8:20~UT) 
is constrained by the 3$\,$mm CARMA data. For the model parameters used to match the NIR data, the
predicted 3$\,$mm flux densities of both flare components (at $\sim$5:30~UT and $\sim$8:20~UT) 
are at or below the flux density limit of 0.1-0.2$\,$Jy, above which no millimeter wavelength
flare event has been detected.
\subsubsection{Expansion modelling of the 26 May flare emission}
\label{subsubsection:adiab-modelling26May2007}
Modelling the 26 May 2008 flare (Fig. \ref{fig:model26}) requires a satisfactory description of 
an initially bright flare with sharply rising and falling flanks 
followed by a 1.5 hour plateau.
This can only be achieved with several source components each covering
a maximum portion of time of about a 30 minute duration.
A smaller number of source components would require larger source 
component sizes resulting in higher magnetic fields and a 
violation of the 100$\,$GHz flux limit.
Given these difficulties, the number of source components is 
of course not unique and the models proposed in Table~\ref{modeldata}
can only be taken as an example.

The overall flare structure is quite reminiscent of the flare 
that was observed on 3 June 2008 simultaneously using APEX 
and the VLT, as described by \cite{eckart2008submm}, with the exception that the 26 May
flare is more continuous than the 3 June 2008 flare. 
An overall flare structure that starts out with a bright event 
followed by a single or a few fainter events was also reported by
\cite{eckart2006mm,eckart2009}.
These events may relate to a common physical structure. An explanation 
could be a disk structure that is expanded
by differential rotation within the disk.
Alternative explanations may involve a common history in terms of either accretion
or magnetic field instability - possibly also within a jet structure
(see \cite{eckart2008submm,eckart2008nirpol} and Sect. \ref{section:alternative}).
 \section{Alternative short jet model}
\label{section:alternative}
Emission from a jet and an underlying accretion process may 
also account for the spectrum of Sgr$\,$A*, with the plasma in the 
jet starting to become optically thin at longer wavelengths with increasing 
distance from the centre. Thus, radiation at different radio wavelengths 
probe different sections of the jet, leading to a correlation with emission 
in other wavelength regimes. Very close to the SMBH, it is difficult
to distinguish between emission from the jet and the accretion flow. 
For this jet model, flux variations may then be caused by accretion 
processes or jet instabilities rather than be the result of modulation 
from an orbiting spot, followed by an adiabatic expansion of jet components.
Details of a compact, weak jet structure are discussed in \cite{markoff2007} (also \citealt{markoff2005}). 

%______________________________________________________________
\section{Summary}
We have presented results from global coordinated multiwavelength observational 
campaigns carried out in 2007 and 2008 using NACO at VLT in the NIR K- and L'-bands, 
and CARMA, ATCA, and IRAM in the mm regime.
%xxxxxxxxxxx

We have presented a new method to obtain concatenated light curves of the compact mm-source Sgr$\,$A*
from single dish telescopes and interferometers in the presence of significant 
flux density contributions from an extended and only partially resolved source.
The method requires several consecutive datasets to be available for each 
participating observatory.
From these data, we have evidence of four different 
flaring events in the NIR, with three of these events being covered later in the mm regime.

We have modelled the flare emission with a model involving adiabatic expansion of synchrotron source 
components, and obtained spectral index values, expansion velocities, and other parameters that are consistent with 
previously published variability studies. The mm-flares were found to occur on average about 1.5$\pm$0.5 hours after 
the NIR flare emission, and the expansion velocities for the various flares ranged from 
0.005$c$~-~0.017$c$. 
%xxxxx

Modelling the NIR-flares with synchrotron components that become
optically thick in the sub-mm wavelength regime and obey the flux
density values or limits in the mm- and X-ray domain
constrains their turnover frequency and flux density.
For the described global experiment, the low mm-regime flux density
limits imply that the turnover frequencies are not lower than about
1.0$\pm$0.3$\,$THz and that the optically thick peak flux densities
at or below this turnover frequencies do not exceed about $\sim$1$\,$Jy
on average.
However, in other experiments the source showed higher flux density
flare emission in the mm-domain (e.g., 0.5-1.5$\,$Jy in the sub-mm;
Yusef-Zadeh et al. 2009, Marrone et al. 2008, Eckart et al. 2008, 2006b)

Further monitoring of the variability of Sgr$\,$A* in different wavelength regimes (including X-ray, NIR, sub-mm, 
and mm radio) and in polarised NIR/radio emission is required to improve our understanding of the adiabatic expansion 
model and improve statistics describing the flaring activity. Current mm-interferometers such as CARMA, PdBI, and ATCA and future 
telescopes such as ALMA which are able to distinguish between emission from Sgr$\,$A* and the thermal emission of the CND and mini-spiral 
surrounding the SMBH, and NIR telescopes with large apertures such as VLT, Keck, and LBT, which can distinguish Sgr$\,$A* from surrounding stars, 
may be combined to provide us with high quality data to study the evolution of synchrotron components in greater detail. 
With future mm-VLBI measurements at frequencies of 230$\,$GHz and above, imaging of the central region of Sgr$\,$A* may become possible, 
enabling a deeper study of the accretion physics and testing of current theories of emission in the region \citep{doeleman2008,doeleman2009}. 

\section*{Acknowledgements}
Part of this work was supported by the German
\emph{Deut\-sche For\-schungs\-ge\-mein\-schaft, DFG\/} via grant SFB 494.
M. Zamaninasab, D. Kunneriath, and R.-S. Lu,
 are members of the International Max Planck Research School (IMPRS) for 
Astronomy and Astrophysics at the MPIfR and the Universities of 
Bonn and Cologne. RS acknowledges support by the Ram\'on y Cajal
programme by the Ministerio de Ciencia y Innovaci\'on of the
government of Spain. Macarena Garcia-Marin is supported by the German federal department for 
education and research (BMBF) under the project numbers: 50OS0502 \& 
50OS0801. N. Sabha is a member of the Bonn Cologne Graduate School (BCGS) of Physics and Astronomy. 

%\vspace*{50mm}
\section*{Appendix:NIR light curves}
\begin{figure}[!Htbp]
\begin{center}
\includegraphics[width=9cm, angle=0]{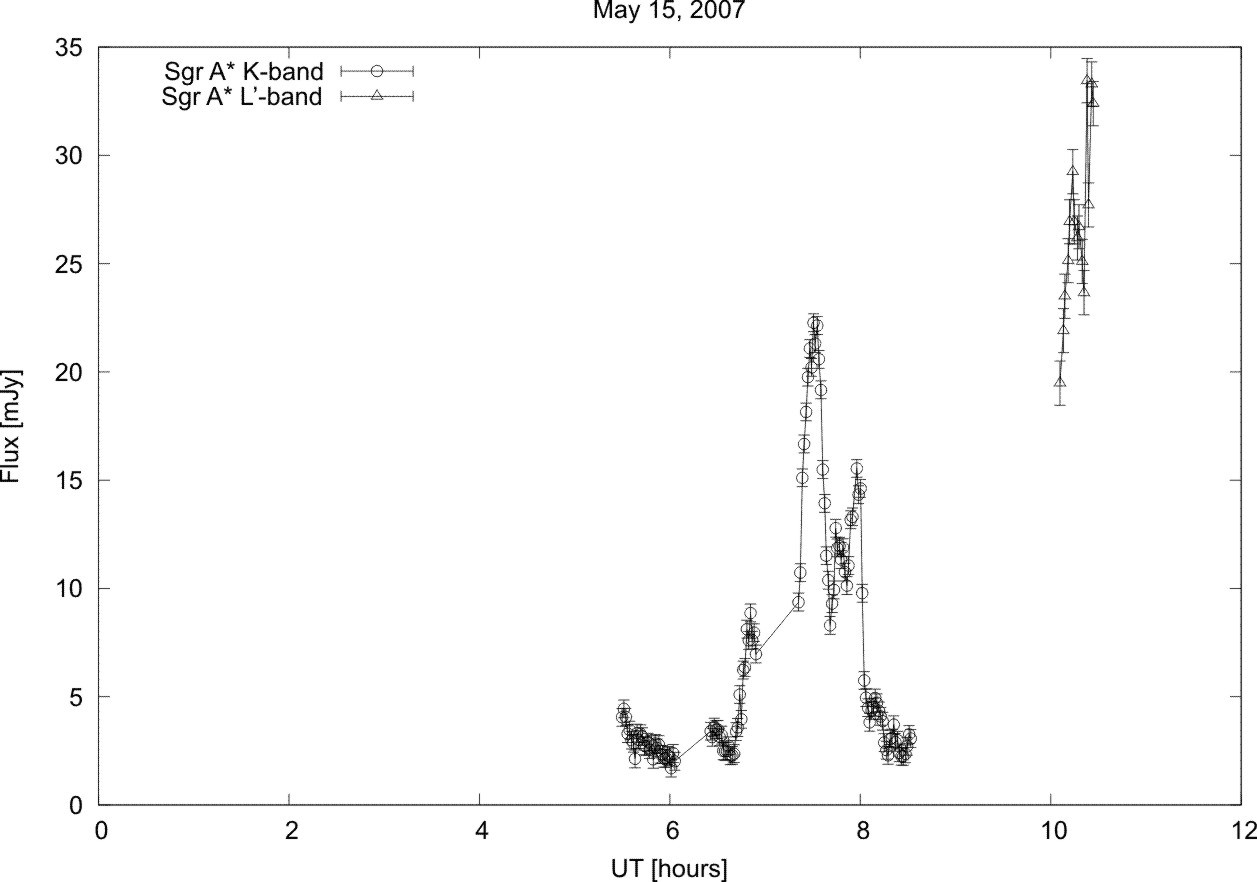}
\end{center}
\caption{Observing time--K-band: 05:29:55 to 08:31:48, L'-band: 10:05:48 to 10:26:45}
\label{fig:1x}
\end{figure}

\begin{figure}[!Htbp]
\begin{center}
\includegraphics[width=9cm, angle=0]{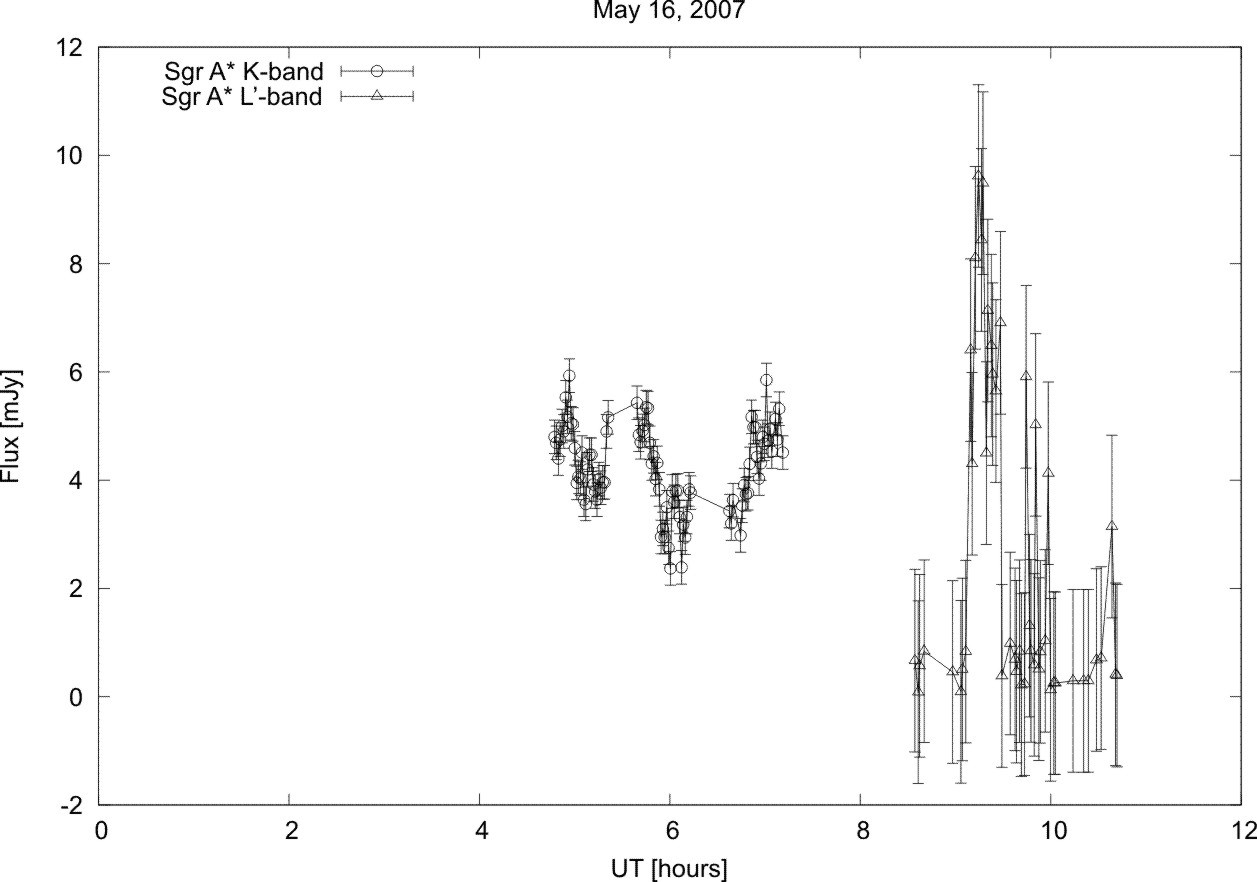}
\end{center}
\caption{Observing time--K-band: 04:47:22 to 07:54:41, L'-band: 08:34:27 to 10:41:46}
\label{fig:2x}
\end{figure}

\begin{figure}[!Htbp]
\begin{center}
\includegraphics[width=9cm, angle=0]{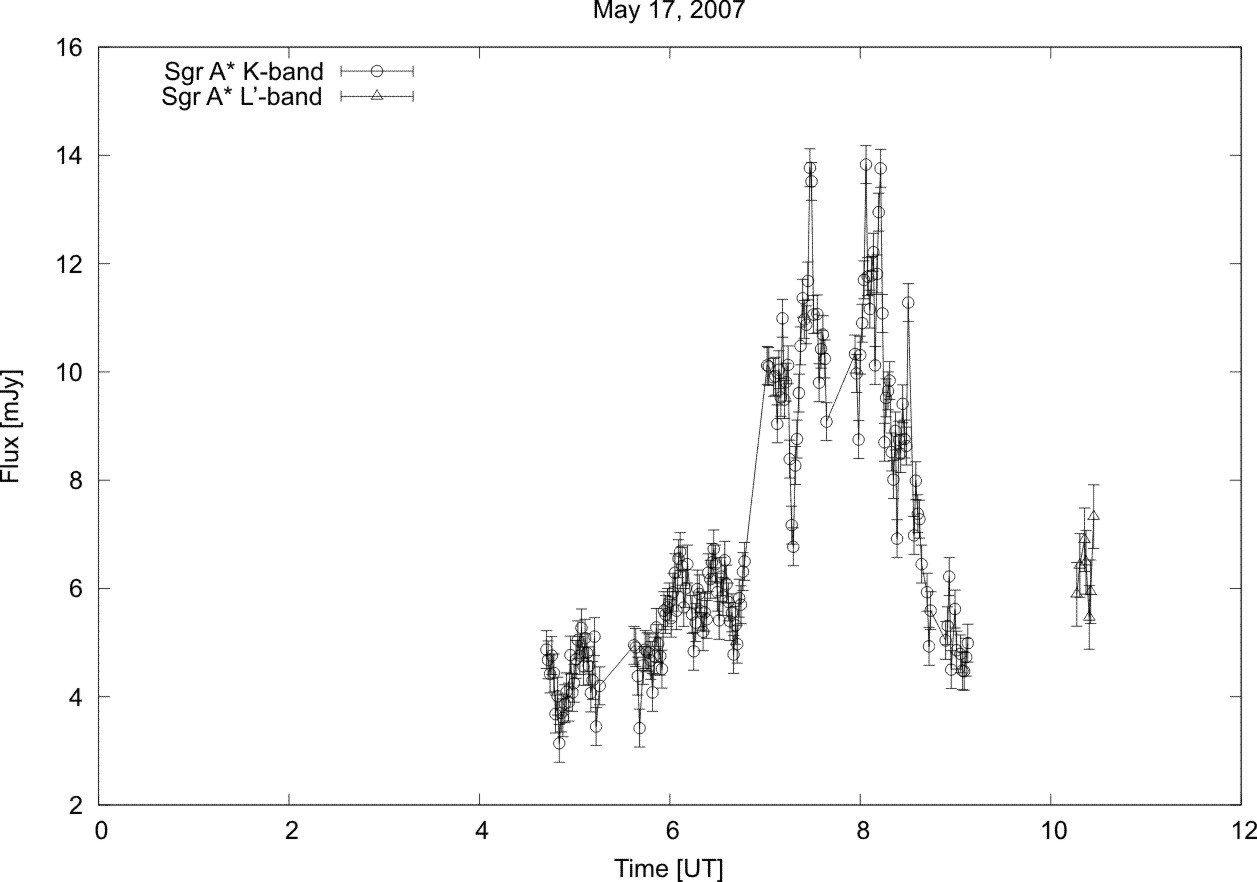}
\end{center}
\caption{Observing time--K-band: 04:42:14 to 09:34:40, L'-band: 10:16:24 to 10:27:06}
\label{fig:3x}
\end{figure}

\begin{figure}[!Htbp]
\begin{center}
\includegraphics[width=9cm, angle=0]{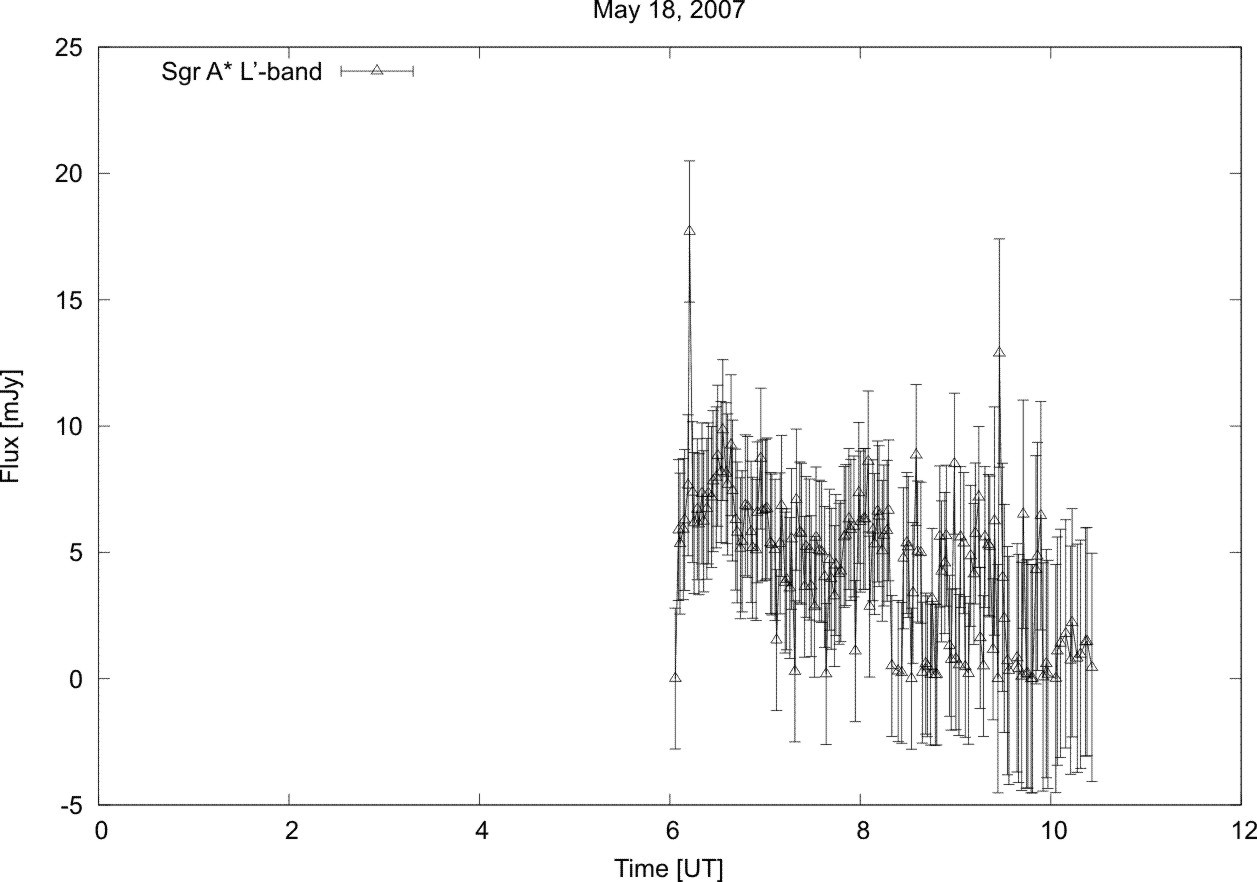}
\end{center}
\caption{Observing time--L'-band: 06:03:26 to 10:26:00}
\label{fig:4x}
\end{figure}

\begin{figure}[!Htbp]
\begin{center}
\includegraphics[width=9cm, angle=0]{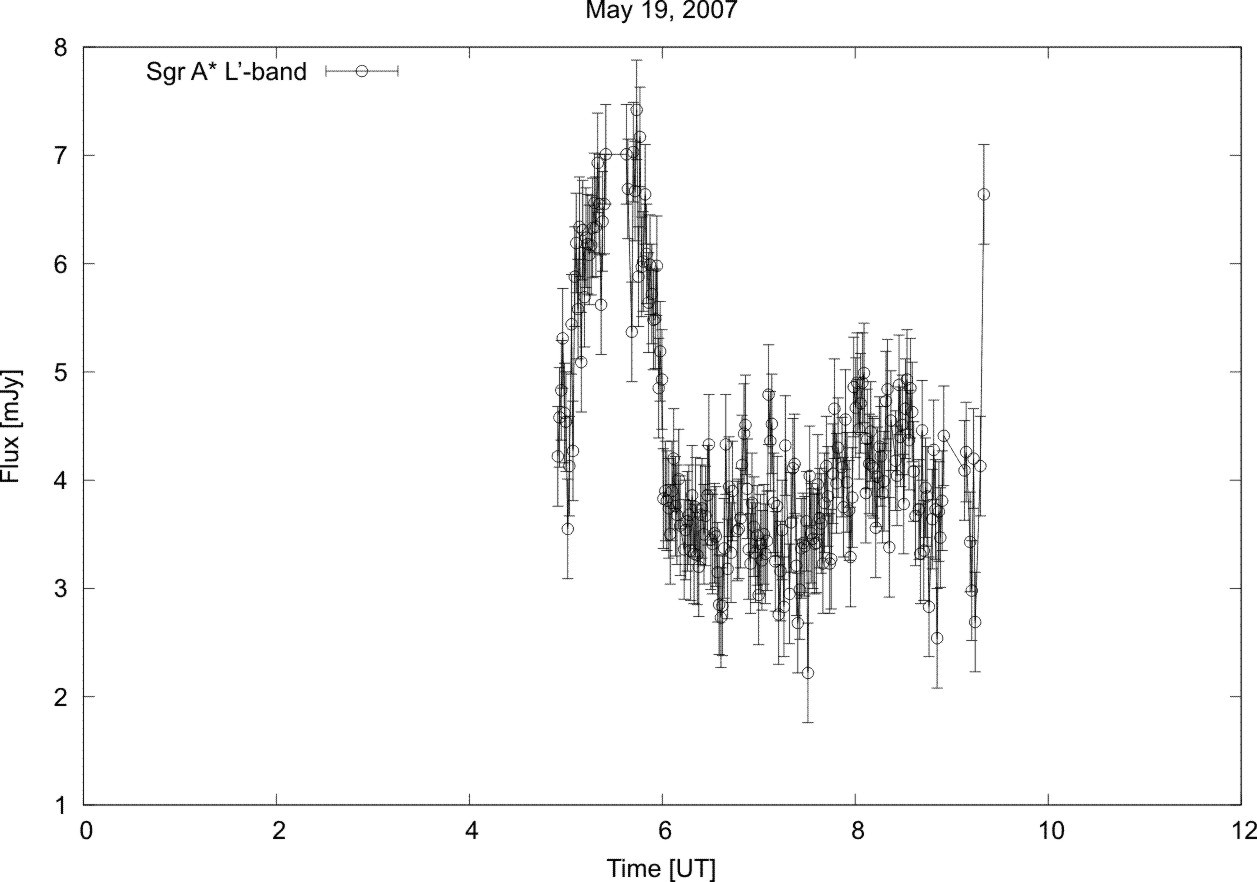}
\end{center}
\caption{Observing time--K-band: 04:55:00 to 09:28:22}
\label{fig:5x}
\end{figure}

The individual NIR K and L'-band light curves from Fig. \ref{fig:1} are shown in Figs. \ref{fig:1x} to \ref{fig:5x}. A spectral index of -0.9 (as we preferentially used in our modelling, see Table \ref{modeldata} and Eckart et al. 2008b, 2009, Yusef-Zadeh et al. 2008) gives us a scaling factor of 0.6, which we use to scale the L'-band data, and then combine the K and L'-band data as shown in Fig. \ref{fig:1} and Figs. \ref{fig:1x} to \ref{fig:5x}. In the remaining figures, no scaling has been applied. The uncertainties in flux measurements were obtained from the reference star S2, which has known flux densities of 22$\pm$1$\,$mJy and 9$\pm$1$\,$mJy in the K and L'-bands, respectively.  

%\vspace*{130mm}

\vspace*{50mm}

\bibliography{13613}{}

\begin{thebibliography}{60}
\expandafter\ifx\csname natexlab\endcsname\relax\def\natexlab#1{#1}\fi

\bibitem[{{Armitage} \& {Reynolds}(2003)}]{armitage2003}
{Armitage}, P.~J. \& {Reynolds}, C.~S. 2003, \mnras, 341, 1041

\bibitem[{{Bock} {et~al.}(2006){Bock}, {Bolatto}, {Hawkins}, {Kemball}, {Lamb},
  {Plambeck}, {Pound}, {Scott}, {Woody}, \& {Wright}}]{carma2006}
{Bock}, D., {Bolatto}, A.~D., {Hawkins}, D.~W., {et~al.} 2006, in Society of
  Photo-Optical Instrumentation Engineers (SPIE) Conference Series, Vol. 6267,
  Society of Photo-Optical Instrumentation Engineers (SPIE) Conference Series

\bibitem[{{Bower} {et~al.}(2002){Bower}, {Falcke}, {Sault}, \&
  {Backer}}]{bower2002}
{Bower}, G.~C., {Falcke}, H., {Sault}, R.~J., \& {Backer}, D.~C. 2002, \apj,
  571, 843

\bibitem[{{Christopher} {et~al.}(2005){Christopher}, {Scoville}, {Stolovy}, \&
  {Yun}}]{christopher2005}
{Christopher}, M.~H., {Scoville}, N.~Z., {Stolovy}, S.~R., \& {Yun}, M.~S.
  2005, \apj, 622, 346

\bibitem[{{Diolaiti} {et~al.}(2000){Diolaiti}, {Bendinelli}, {Bonaccini},
  {Close}, {Currie}, \& {Parmeggiani}}]{diolaiti2000}
{Diolaiti}, E., {Bendinelli}, O., {Bonaccini}, D., {et~al.} 2000, in Society of
  Photo-Optical Instrumentation Engineers (SPIE) Conference Series, Vol. 4007,
  Society of Photo-Optical Instrumentation Engineers (SPIE) Conference Series,
  ed. {P.~L.~Wizinowich}, 879--888

\bibitem[{{Do} {et~al.}(2008){Do}, {Ghez}, {Morris}, {Yelda}, {Lu},
  {Hornstein}, \& {Matthews}}]{do2008}
{Do}, T., {Ghez}, A.~M., {Morris}, M.~R., {et~al.} 2008, Journal of Physics
  Conference Series, 131, 012003

\bibitem[{{Do} {et~al.}(2009){Do}, {Ghez}, {Morris}, {Yelda}, {Meyer}, {Lu},
  {Hornstein}, \& {Matthews}}]{do2009}
{Do}, T., {Ghez}, A.~M., {Morris}, M.~R., {et~al.} 2009, \apj, 691, 1021

\bibitem[{{Doeleman} {et~al.}(2009){Doeleman}, {Fish}, {Broderick}, {Loeb}, \&
  {Rogers}}]{doeleman2009}
{Doeleman}, S.~S., {Fish}, V.~L., {Broderick}, A.~E., {Loeb}, A., \& {Rogers},
  A.~E.~E. 2009, \apj, 695, 59

\bibitem[{{Doeleman} {et~al.}(2008){Doeleman}, {Weintroub}, {Rogers},
  {Plambeck}, {Freund}, {Tilanus}, {Friberg}, {Ziurys}, {Moran}, {Corey},
  {Young}, {Smythe}, {Titus}, {Marrone}, {Cappallo}, {Bock}, {Bower},
  {Chamberlin}, {Davis}, {Krichbaum}, {Lamb}, {Maness}, {Niell}, {Roy},
  {Strittmatter}, {Werthimer}, {Whitney}, \& {Woody}}]{doeleman2008}
{Doeleman}, S.~S., {Weintroub}, J., {Rogers}, A.~E.~E., {et~al.} 2008, \nat,
  455, 78

\bibitem[{{Dov{\v c}iak} {et~al.}(2004){Dov{\v c}iak}, {Karas}, \&
  {Yaqoob}}]{dovciak2004}
{Dov{\v c}iak}, M., {Karas}, V., \& {Yaqoob}, T. 2004, \apjs, 153, 205

\bibitem[{{Eckart} {et~al.}(2004){Eckart}, {Baganoff}, {Morris}, {Bautz},
  {Brandt}, {Garmire}, {Genzel}, {Ott}, {Ricker}, {Straubmeier}, {Viehmann},
  {Sch{\"o}del}, {Bower}, \& {Goldston}}]{eckart2004}
{Eckart}, A., {Baganoff}, F.~K., {Morris}, M., {et~al.} 2004, \aap, 427, 1

\bibitem[{{Eckart} {et~al.}(2009){Eckart}, {Baganoff}, {Morris}, {Kunneriath},
  {Zamaninasab}, {Witzel}, {Sch{\"o}del}, {Garc{\'{\i}}a-Mar{\'{\i}}n},
  {Meyer}, {Bower}, {Marrone}, {Bautz}, {Brandt}, {Garmire}, {Ricker},
  {Straubmeier}, {Roberts}, {Muzic}, {Mauerhan}, \& {Zensus}}]{eckart2009}
{Eckart}, A., {Baganoff}, F.~K., {Morris}, M.~R., {et~al.} 2009, \aap, 500, 935

\bibitem[{{Eckart} {et~al.}(2006{\natexlab{a}}){Eckart}, {Baganoff},
  {Sch{\"o}del}, {Morris}, {Genzel}, {Bower}, {Marrone}, {Moran}, {Viehmann},
  {Bautz}, {Brandt}, {Garmire}, {Ott}, {Trippe}, {Ricker}, {Straubmeier},
  {Roberts}, {Yusef-Zadeh}, {Zhao}, \& {Rao}}]{eckart2006mm}
{Eckart}, A., {Baganoff}, F.~K., {Sch{\"o}del}, R., {et~al.}
  2006{\natexlab{a}}, \aap, 450, 535

\bibitem[{{Eckart} {et~al.}(2008{\natexlab{a}}){Eckart}, {Baganoff},
  {Zamaninasab}, {Morris}, {Sch{\"o}del}, {Meyer}, {Muzic}, {Bautz}, {Brandt},
  {Garmire}, {Ricker}, {Kunneriath}, {Straubmeier}, {Duschl}, {Dovciak},
  {Karas}, {Markoff}, {Najarro}, {Mauerhan}, {Moultaka}, \&
  {Zensus}}]{eckart2008nirpol}
{Eckart}, A., {Baganoff}, F.~K., {Zamaninasab}, M., {et~al.}
  2008{\natexlab{a}}, \aap, 479, 625

\bibitem[{{Eckart} \& {Genzel}(1996)}]{eckart1996}
{Eckart}, A. \& {Genzel}, R. 1996, \nat, 383, 415

\bibitem[{{Eckart} {et~al.}(2002){Eckart}, {Genzel}, {Ott}, \&
  {Sch{\"o}del}}]{eckart2002}
{Eckart}, A., {Genzel}, R., {Ott}, T., \& {Sch{\"o}del}, R. 2002, \mnras, 331,
  917

\bibitem[{{Eckart} {et~al.}(2008{\natexlab{b}}){Eckart}, {Sch{\"o}del},
  {Garc{\'{\i}}a-Mar{\'{\i}}n}, {Witzel}, {Weiss}, {Baganoff}, {Morris},
  {Bertram}, {Dov{\v c}iak}, {Duschl}, {Karas}, {K{\"o}nig}, {Krichbaum},
  {Krips}, {Kunneriath}, {Lu}, {Markoff}, {Mauerhan}, {Meyer}, {Moultaka},
  {Mu{\v z}i{\'c}}, {Najarro}, {Pott}, {Schuster}, {Sjouwerman}, {Straubmeier},
  {Thum}, {Vogel}, {Wiesemeyer}, {Zamaninasab}, \& {Zensus}}]{eckart2008submm}
{Eckart}, A., {Sch{\"o}del}, R., {Garc{\'{\i}}a-Mar{\'{\i}}n}, M., {et~al.}
  2008{\natexlab{b}}, \aap, 492, 337

\bibitem[{{Eckart} {et~al.}(2006{\natexlab{b}}){Eckart}, {Sch{\"o}del},
  {Meyer}, {Trippe}, {Ott}, \& {Genzel}}]{eckart2006nirpol}
{Eckart}, A., {Sch{\"o}del}, R., {Meyer}, L., {et~al.} 2006{\natexlab{b}},
  \aap, 455, 1

\bibitem[{{Eisenhauer} {et~al.}(2005){Eisenhauer}, {Genzel}, {Alexander},
  {Abuter}, {Paumard}, {Ott}, {Gilbert}, {Gillessen}, {Horrobin}, {Trippe},
  {Bonnet}, {Dumas}, {Hubin}, {Kaufer}, {Kissler-Patig}, {Monnet},
  {Str{\"o}bele}, {Szeifert}, {Eckart}, {Sch{\"o}del}, \&
  {Zucker}}]{eisenhauer2005}
{Eisenhauer}, F., {Genzel}, R., {Alexander}, T., {et~al.} 2005, \apj, 628, 246

\bibitem[{{Eisenhauer} {et~al.}(2003){Eisenhauer}, {Sch{\"o}del}, {Genzel},
  {Ott}, {Tecza}, {Abuter}, {Eckart}, \& {Alexander}}]{eisenhauer2003}
{Eisenhauer}, F., {Sch{\"o}del}, R., {Genzel}, R., {et~al.} 2003, \apjl, 597,
  L121

\bibitem[{{Genzel} {et~al.}(1997){Genzel}, {Eckart}, {Ott}, \&
  {Eisenhauer}}]{genzel1997}
{Genzel}, R., {Eckart}, A., {Ott}, T., \& {Eisenhauer}, F. 1997, \mnras, 291,
  219

\bibitem[{{Genzel} {et~al.}(2000){Genzel}, {Pichon}, {Eckart}, {Gerhard}, \&
  {Ott}}]{genzel2000}
{Genzel}, R., {Pichon}, C., {Eckart}, A., {Gerhard}, O.~E., \& {Ott}, T. 2000,
  \mnras, 317, 348

\bibitem[{{Ghez} {et~al.}(2009){Ghez}, {Morris}, {Lu}, {Weinberg}, {Matthews},
  {Alexander}, {Armitage}, {Becklin}, {Brown}, {Campbell}, {Do}, {Eckart},
  {Genzel}, {Gould}, {Hansen}, {Ho}, {Lo}, {Loeb}, {Melia}, {Merritt},
  {Milosavljevic}, {Perets}, {Rasio}, {Reid}, {Salim}, {Sch{\"o}del}, \&
  {Yelda}}]{ghez2009}
{Ghez}, A., {Morris}, M., {Lu}, J., {et~al.} 2009, in Astronomy, Vol. 2010, AGB
  Stars and Related Phenomenastro2010: The Astronomy and Astrophysics Decadal
  Survey, 89--+

\bibitem[{{Ghez} {et~al.}(2004{\natexlab{a}}){Ghez}, {Hornstein}, {Bouchez},
  {Le Mignant}, {Lu}, {Matthews}, {Morris}, {Wizinowich}, \&
  {Becklin}}]{ghez2004b}
{Ghez}, A.~M., {Hornstein}, S.~D., {Bouchez}, A., {et~al.} 2004{\natexlab{a}},
  in Bulletin of the American Astronomical Society, Vol.~36, Bulletin of the
  American Astronomical Society, 1384--+

\bibitem[{{Ghez} {et~al.}(1998){Ghez}, {Klein}, {Morris}, \&
  {Becklin}}]{ghez1998}
{Ghez}, A.~M., {Klein}, B.~L., {Morris}, M., \& {Becklin}, E.~E. 1998, \apj,
  509, 678

\bibitem[{{Ghez} {et~al.}(2000){Ghez}, {Morris}, {Becklin}, {Tanner}, \&
  {Kremenek}}]{ghez2000}
{Ghez}, A.~M., {Morris}, M., {Becklin}, E.~E., {Tanner}, A., \& {Kremenek}, T.
  2000, \nat, 407, 349

\bibitem[{{Ghez} {et~al.}(2005){Ghez}, {Salim}, {Hornstein}, {Tanner}, {Lu},
  {Morris}, {Becklin}, \& {Duch{\^e}ne}}]{ghez2005}
{Ghez}, A.~M., {Salim}, S., {Hornstein}, S.~D., {et~al.} 2005, \apj, 620, 744

\bibitem[{{Ghez} {et~al.}(2004{\natexlab{b}}){Ghez}, {Wright}, {Matthews},
  {Thompson}, {Le Mignant}, {Tanner}, {Hornstein}, {Morris}, {Becklin}, \&
  {Soifer}}]{ghez2004a}
{Ghez}, A.~M., {Wright}, S.~A., {Matthews}, K., {et~al.} 2004{\natexlab{b}},
  \apjl, 601, L159

\bibitem[{{Gillessen} {et~al.}(2009){Gillessen}, {Eisenhauer}, {Trippe},
  {Alexander}, {Genzel}, {Martins}, \& {Ott}}]{gillessen2009}
{Gillessen}, S., {Eisenhauer}, F., {Trippe}, S., {et~al.} 2009, \apj, 692, 1075

\bibitem[{{Gould}(1979)}]{gould1979}
{Gould}, R.~J. 1979, \aap, 76, 306

\bibitem[{{Guesten} {et~al.}(1987){Guesten}, {Genzel}, {Wright}, {Jaffe},
  {Stutzki}, \& {Harris}}]{guesten1987}
{Guesten}, R., {Genzel}, R., {Wright}, M.~C.~H., {et~al.} 1987, \apj, 318, 124

\bibitem[{{Herrnstein} {et~al.}(2004){Herrnstein}, {Zhao}, {Bower}, \&
  {Goss}}]{herrnstein2004}
{Herrnstein}, R.~M., {Zhao}, J., {Bower}, G.~C., \& {Goss}, W.~M. 2004, \aj,
  127, 3399

\bibitem[{{Kunneriath} {et~al.}(2008){Kunneriath}, {Eckart}, {Vogel},
  {Sjouwerman}, {Wiesemeyer}, {Sch{\"o}del}, {Baganoff}, {Morris}, {Bertram},
  {Dovciak}, {Dowries}, {Duschl}, {Karas}, {Konig}, {Krichbaum}, {Krips}, {Lu},
  {Markoff}, {Mauerhan}, {Meyer}, {Moultaka}, {Muzic}, {Najarro}, {Schuster},
  {Straubmeier}, {Thum}, {Witzel}, {Zamaninasab}, \& {Zensus}}]{kunneriath2008}
{Kunneriath}, D., {Eckart}, A., {Vogel}, S., {et~al.} 2008, Journal of Physics
  Conference Series, 131, 012006

\bibitem[{{Lenzen} {et~al.}(2003){Lenzen}, {Hartung}, {Brandner}, {Finger},
  {Hubin}, {Lacombe}, {Lagrange}, {Lehnert}, {Moorwood}, \&
  {Mouillet}}]{lenzen2003}
{Lenzen}, R., {Hartung}, M., {Brandner}, W., {et~al.} 2003, in Society of
  Photo-Optical Instrumentation Engineers (SPIE) Conference Series, Vol. 4841,
  Society of Photo-Optical Instrumentation Engineers (SPIE) Conference Series,
  ed. {M.~Iye \& A.~F.~M.~Moorwood}, 944--952

\bibitem[{{Lu} {et~al.}(2008){Lu}, {Krichbaum}, {Eckart}, {K{\"o}nig},
  {Kunneriath}, {Witzel}, {Witzel}, \& {Zensus}}]{lu2008}
{Lu}, R., {Krichbaum}, T.~P., {Eckart}, A., {et~al.} 2008, Journal of Physics
  Conference Series, 131, 012059

\bibitem[{{Lu} {et~al.}(2009){Lu}, {Krichbaum}, {Eckart}, {K{\"o}nig},
  {Kunneriath}, {Witzel}, {Witzel}, \& {Zensus}}]{lu2009}
{Lu}, R., {Krichbaum}, T.~P., {Eckart}, A., {et~al.} 2009, \aap, in prep.

\bibitem[{{Markoff} {et~al.}(2007){Markoff}, {Bower}, \&
  {Falcke}}]{markoff2007}
{Markoff}, S., {Bower}, G.~C., \& {Falcke}, H. 2007, \mnras, 379, 1519

\bibitem[{{Markoff} {et~al.}(2001){Markoff}, {Falcke}, {Yuan}, \&
  {Biermann}}]{markoff2001}
{Markoff}, S., {Falcke}, H., {Yuan}, F., \& {Biermann}, P.~L. 2001, \aap, 379,
  L13

\bibitem[{{Markoff} {et~al.}(2005){Markoff}, {Nowak}, \& {Wilms}}]{markoff2005}
{Markoff}, S., {Nowak}, M.~A., \& {Wilms}, J. 2005, \apj, 635, 1203

\bibitem[{{Marrone} {et~al.}(2008){Marrone}, {Baganoff}, {Morris}, {Moran},
  {Ghez}, {Hornstein}, {Dowell}, {Mu{\~n}oz}, {Bautz}, {Ricker}, {Brandt},
  {Garmire}, {Lu}, {Matthews}, {Zhao}, {Rao}, \& {Bower}}]{marrone2008}
{Marrone}, D.~P., {Baganoff}, F.~K., {Morris}, M.~R., {et~al.} 2008, \apj, 682,
  373

\bibitem[{{Marscher}(1983)}]{marscher1983}
{Marscher}, A.~P. 1983, \apj, 264, 296

\bibitem[{{Mauerhan} {et~al.}(2005){Mauerhan}, {Morris}, {Walter}, \&
  {Baganoff}}]{mauerhan2005}
{Mauerhan}, J.~C., {Morris}, M., {Walter}, F., \& {Baganoff}, F.~K. 2005,
  \apjl, 623, L25

\bibitem[{{Meyer} {et~al.}(2006{\natexlab{a}}){Meyer}, {Eckart}, {Sch{\"o}del},
  {Duschl}, {Mu{\v z}i{\'c}}, {Dov{\v c}iak}, \& {Karas}}]{meyer2006b}
{Meyer}, L., {Eckart}, A., {Sch{\"o}del}, R., {et~al.} 2006{\natexlab{a}},
  \aap, 460, 15

\bibitem[{{Meyer} {et~al.}(2007){Meyer}, {Sch{\"o}del}, {Eckart}, {Duschl},
  {Karas}, \& {Dov{\v c}iak}}]{meyer2007}
{Meyer}, L., {Sch{\"o}del}, R., {Eckart}, A., {et~al.} 2007, \aap, 473, 707

\bibitem[{{Meyer} {et~al.}(2006{\natexlab{b}}){Meyer}, {Sch{\"o}del}, {Eckart},
  {Karas}, {Dov{\v c}iak}, \& {Duschl}}]{meyer2006a}
{Meyer}, L., {Sch{\"o}del}, R., {Eckart}, A., {et~al.} 2006{\natexlab{b}},
  \aap, 458, L25

\bibitem[{{Pech{\'a}{\v c}ek} {et~al.}(2008){Pech{\'a}{\v c}ek}, {Karas}, \&
  {Czerny}}]{pechacek2008}
{Pech{\'a}{\v c}ek}, T., {Karas}, V., \& {Czerny}, B. 2008, \aap, 487, 815

\bibitem[{{Rousset} {et~al.}(2003){Rousset}, {Lacombe}, {Puget}, {Hubin},
  {Gendron}, {Fusco}, {Arsenault}, {Charton}, {Feautrier}, {Gigan}, {Kern},
  {Lagrange}, {Madec}, {Mouillet}, {Rabaud}, {Rabou}, {Stadler}, \&
  {Zins}}]{rousset2003}
{Rousset}, G., {Lacombe}, F., {Puget}, P., {et~al.} 2003, in Society of
  Photo-Optical Instrumentation Engineers (SPIE) Conference Series, Vol. 4839,
  Society of Photo-Optical Instrumentation Engineers (SPIE) Conference Series,
  ed. {P.~L.~Wizinowich \& D.~Bonaccini}, 140--149

\bibitem[{{Sabha} {et~al.}(2010){Sabha}, {Witzel}, {Eckart}, {Buchholz},
  {Bremer}, {Giessuebel}, {Garcia-Marin}, {Kunneriath}, {Muzic}, {Schoedel},
  {Straubmeier}, {Zamaninasab}, \& {Zernickel}}]{sabha2010}
{Sabha}, N., {Witzel}, G., {Eckart}, A., {et~al.} 2010, \aap, in press

\bibitem[{{Sch{\"o}del} {et~al.}(2003){Sch{\"o}del}, {Ott}, {Genzel}, {Eckart},
  {Mouawad}, \& {Alexander}}]{schoedel2003}
{Sch{\"o}del}, R., {Ott}, T., {Genzel}, R., {et~al.} 2003, \apj, 596, 1015

\bibitem[{{Sch{\"o}del} {et~al.}(2002){Sch{\"o}del}, {Ott}, {Genzel},
  {Hofmann}, {Lehnert}, {Eckart}, {Mouawad}, {Alexander}, {Reid}, {Lenzen},
  {Hartung}, {Lacombe}, {Rouan}, {Gendron}, {Rousset}, {Lagrange}, {Brandner},
  {Ageorges}, {Lidman}, {Moorwood}, {Spyromilio}, {Hubin}, \&
  {Menten}}]{schoedel2002}
{Sch{\"o}del}, R., {Ott}, T., {Genzel}, R., {et~al.} 2002, \nat, 419, 694

\bibitem[{{van der Laan}(1966)}]{laan1966}
{van der Laan}, H. 1966, \nat, 211, 1131

\bibitem[{{Yusef-Zadeh} {et~al.}(2006{\natexlab{a}}){Yusef-Zadeh}, {Bushouse},
  {Dowell}, {Wardle}, {Roberts}, {Heinke}, {Bower}, {Vila-Vilar{\'o}},
  {Shapiro}, {Goldwurm}, \& {B{\'e}langer}}]{zadeh2006a}
{Yusef-Zadeh}, F., {Bushouse}, H., {Dowell}, C.~D., {et~al.}
  2006{\natexlab{a}}, \apj, 644, 198

\bibitem[{{Yusef-Zadeh} {et~al.}(2009){Yusef-Zadeh}, {Bushouse}, {Wardle},
  {Heinke}, {Roberts}, {Dowell}, {Brunthaler}, {Reid}, {Martin}, {Marrone},
  {Porquet}, {Grosso}, {Dodds-Eden}, {Bower}, {Wiesemeyer}, {Miyazaki}, {Pal},
  {Gillessen}, {Goldwurm}, {Trap}, \& {Maness}}]{zadeh2009}
{Yusef-Zadeh}, F., {Bushouse}, H., {Wardle}, M., {et~al.} 2009, ArXiv e-prints

\bibitem[{{Yusef-Zadeh} {et~al.}(2006{\natexlab{b}}){Yusef-Zadeh}, {Roberts},
  {Wardle}, {Heinke}, \& {Bower}}]{zadeh2006b}
{Yusef-Zadeh}, F., {Roberts}, D., {Wardle}, M., {Heinke}, C.~O., \& {Bower},
  G.~C. 2006{\natexlab{b}}, \apj, 650, 189

\bibitem[{{Yusef-Zadeh} {et~al.}(2008){Yusef-Zadeh}, {Wardle}, {Heinke},
  {Dowell}, {Roberts}, {Baganoff}, \& {Cotton}}]{zadeh2008}
{Yusef-Zadeh}, F., {Wardle}, M., {Heinke}, C., {et~al.} 2008, \apj, 682, 361

\bibitem[{{Zamaninasab} {et~al.}(2008{\natexlab{a}}){Zamaninasab}, {Eckart},
  {Kunneriath}, {Witzel}, {Sch{\"o}del}, {Meyer}, {Dovciak}, {Karas},
  {K{\"o}nig}, {Krichbaum}, {Lu}, {Straubmeier}, \& {Zensus}}]{zamani2008b}
{Zamaninasab}, M., {Eckart}, A., {Kunneriath}, D., {et~al.} 2008{\natexlab{a}},
  Memorie della Societa Astronomica Italiana, 79, 1054

\bibitem[{{Zamaninasab} {et~al.}(2008{\natexlab{b}}){Zamaninasab}, {Eckart},
  {Meyer}, {Sch{\"o}del}, {Dovciak}, {Karas}, {Kunneriath}, {Witzel},
  {Gie{\ss}{\"u}bel}, {K{\"o}nig}, {Straubmeier}, \& {Zensus}}]{zamani2008a}
{Zamaninasab}, M., {Eckart}, A., {Meyer}, L., {et~al.} 2008{\natexlab{b}},
  Journal of Physics Conference Series, 131, 012008

\bibitem[{{Zamaninasab} {et~al.}(2010){Zamaninasab}, {Eckart}, {Witzel},
  {Dovciak}, {Karas}, {Sch{\"o}del}, {Gie{\ss}{\"u}bel}, {Bremer},
  {Garc{\'{\i}}a-Mar{\'{\i}}n}, {Kunneriath}, {Mu{\v z}i{\'c}}, {Nishiyama},
  {Sabha}, {Straubmeier}, \& {Zensus}}]{zamani2010}
{Zamaninasab}, M., {Eckart}, A., {Witzel}, G., {et~al.} 2010, \aap, 510, A3+

\bibitem[{{Zhao} {et~al.}(2004){Zhao}, {Herrnstein}, {Bower}, {Goss}, \&
  {Liu}}]{zhao2004}
{Zhao}, J., {Herrnstein}, R.~M., {Bower}, G.~C., {Goss}, W.~M., \& {Liu}, S.~M.
  2004, \apjl, 603, L85

\bibitem[{{Zhao} {et~al.}(2003){Zhao}, {Young}, {Herrnstein}, {Ho}, {Tsutsumi},
  {Lo}, {Goss}, \& {Bower}}]{zhao2003}
{Zhao}, J., {Young}, K.~H., {Herrnstein}, R.~M., {et~al.} 2003, \apjl, 586, L29

\end{thebibliography}
\bibliographystyle{aa}

\end{document}